\documentclass[twoside]{article}%
\usepackage[utf8x]{inputenc}
\usepackage{amsmath, amssymb}
\usepackage[left=2cm, right=2cm]{geometry}
\usepackage{bbm}
\usepackage{multicol}
\usepackage{enumitem}
\usepackage{braket}
\usepackage{amsmath}
\usepackage{amsfonts}
\usepackage{amssymb}
\usepackage{graphicx}%
\newcommand{\n}{\nonumber\\}
\DeclareMathOperator{\Tr}{Tr}
\begin{document}

\title{A Three-Flavor Chiral Effective Model with Four Baryonic Multiplets within the
Mirror Assignment}
\author{Lisa Olbrich$^{1}$, Mikl\'{o}s Z\'{e}t\'{e}nyi$^{1,2}$, Francesco Giacosa$^{1,3}$, and Dirk H.\ Rischke$^{1}$ \\ 
\emph{{\small $^{1}$Institute for Theoretical Physics, Goethe University }}\\[-0.1cm] 
\emph{{\small Max-von-Laue--Str.\ 1, D-60438 Frankfurt am Main, Germany}}\\[-0.3cm] \\
\emph{{\small $^{2}$ Wigner Research Center for Physics}}\\[-0.1cm]
\emph{{\small Konkoly Thege Mikl\'{o}s \'{u}t 29-33, H-1121 Budapest, Hungary}}\\[-0.3cm] \\
\emph{{\small $^{3}$Institute of Physics, Jan Kochanowski University,}}\\[-0.1cm] 
\emph{{\small ul.\ Swietokrzyska 15, 25-406 Kielce, Poland}}}
\maketitle

\begin{abstract}
In the case of three quark flavors, (pseudo)scalar diquarks transform as antiquarks under chiral transformations.
We construct four spin-1/2 baryonic multiplets from left- and right-handed quarks as well as
left- and right-handed diquarks. The fact that two of these multiplets transform
in a ``mirror'' way allows for chirally invariant mass terms. We then embed these
baryonic multiplets into the Lagrangian of the so-called extended Linear Sigma Model, which
features (pseudo)scalar and (axial-)vector mesons, as well as glueballs.
Reducing the Lagrangian to the two-flavor case, we obtain
four doublets of nucleonic states. These mix to produce four
experimentally observed states with definite parity: the positive-parity nucleon $N(939)$ and
Roper resonance $N(1440)$, as well as the negative-parity resonances $N(1535)$
and $N(1650)$. We determine the parameters of the nucleonic part of the Lagrangian from
a fit to masses and decay properties of the aforementioned states. Studying the limit of vanishing
quark condensate, we conclude that $N(939)$ and $N(1535)$, as well as $N(1440)$ and $N(1650)$
form pairs of chiral partners.
\end{abstract}

\section{Introduction}

The strong interaction determines the masses of the baryons and their interactions with mesons.
At low energies, chiral effective approaches play an important role to describe these 
phenomena \cite{gasioro}. Most notably, one can use chiral perturbation theory, which is
based on the non-linear realization of chiral symmetry
\cite{meissner,jorge,Baru:2010xn}, or $\sigma$-like models, which are
based on the linear realization of chiral symmetry
\cite{koch,lee,dmitrasinovic,dmitra2,Gallas:2009qp,Gallas:2013ipa}.

An effective model based on linearly realized chiral symmetry as well as dilatation invariance
has been constructed in Refs.\ \cite{Gallas:2009qp,Gallas:2013ipa,denis,denisnf3,stani,nf4}. 
This so-called extended Linear Sigma Model (eLSM) also contains
anomalous, explicit, and spontaneous symmetry breaking (SSB) terms in order to reproduce
known features of the strong interaction. The mesonic sector of the eLSM, first developed
for two flavors ($N_{f}=2$) \cite{denis} and further extended to $N_{f}=3$
\cite{denisnf3,stani} and $N_{f}=4$ \cite{nf4}, includes scalar and pseudoscalar as
well as vector and axial-vector degrees of freedom. 
It is able to describe mesonic masses and decays of quark-antiquark mesons up to $1.7$ GeV
within reasonable accuracy [for precursory models including (axial-)vector degrees of freedom
see Ref.\ \cite{ko}].
Moreover, in agreement with results from other approaches \cite{scalars}, the model implies
that the scalar quark-antiquark states are heavier than $1$ GeV and
that $f_{0}(1710)$ is predominantly gluonic \cite{stani}. As a consequence,
the chiral partner of the pion is the resonance $f_{0}(1370)$ and not the
light scalar state $f_{0}(500)$ [which, together with the other light scalar
mesons, is a state made from (at least) four quarks, either a resonance 
dynamically generated in the pseudoscalar scattering continuum
or a diquark-diquark configuration, see e.g.\ Refs.\ \cite{jaffe,tetraquarks,oller,wolkanowskinew,pelaezrev}].

In the standard linear sigma model with nucleons only, chiral symmetry requires that
the mass of the nucleon is (apart from explicit symmetry breaking effects from the current quark masses), 
solely generated by the chiral condensate, $m_{N}$ $\propto\left\langle \bar
{q}q\right\rangle $. However, when one includes the chiral partner of the nucleon,
one can either assume that the partner transforms as the nucleon under chiral transformations
(the so-called ``naive'' assignment), or that it transforms in a ``mirror'' way (the so-called
``mirror'' assignment) \cite{Detar:1988kn,Jido:2001nt,ziesche,sasakimishustin,glozman}. 
The latter one allows for an additional
chirally invariant mass term, which physically parametrizes the contribution to the nucleon
mass that arises from sources other than the chiral condensate (e.g.\ a gluon or a four-quark condensate).
Nucleons and their chiral partners have been studied within the eLSM in the mirror assignment 
in Refs.\ \cite{Gallas:2009qp,Gallas:2013ipa,achim}, indicating
that the contribution to the nucleon mass from these other sources is sizable.

In this work, we extend the work of Refs.\ \cite{Gallas:2009qp,Gallas:2013ipa} to
the case of baryons with $N_f=3$ flavors. This extension will enable us
to address in future work important problems in hadron physics, such as scattering processes 
involving strange hadrons \cite{pptopkaon,pptoppphi,pptoppetaprime,kaonp},
and in astrophysics, e.g.\ the hyperon puzzle for compact stars \cite{hyperonstar,giuseppe}.

For baryons, the extension to the $N_f=3$ case is not as straightforward as for mesons.
In the $N_f=2$ case, the nucleon multiplet is
described by a spinor isodoublet, $\psi_{N}=(p,n)^{T}$, where $p$ and $n$ are 
the proton and the neutron, respectively.
However, in the $N_f=3$ case the $J^{P}=\frac{1}{2}^+$ baryon octet is given by a
$3\times3$ matrix, 
\begin{equation}
\left(
\begin{array}
[c]{ccc}%
\frac{\Lambda}{\sqrt{6}}+\frac{\Sigma^{0}}{\sqrt{2}} & \Sigma^{+} & p\\
\Sigma^{-} & \frac{\Lambda}{\sqrt{6}}-\frac{\Sigma^{0}}{\sqrt{2}} & n\\
\Xi^{-} & \Xi^{0} & -\frac{2\Lambda}{\sqrt{6}}%
\end{array}
\right)  \text{ .} \label{groundstatebaryons}%
\end{equation}

Adding the chiral partner $J^P=\frac{1}{2}^-$ multiplet is also not as straightforward as
in the $N_f=2$ case. Here, we utilize a quark-diquark ``quasi-particle'' picture for the
baryonic substructure. We assume that the diquark is a (pseudo)scalar and lives in the color- and flavor-antitriplet
representation [a so-called ``good'' diquark in the nomenclature of Jaffe \cite{jaffe}], 
such that it transforms as an antiquark. Then, it is quite natural that
$\mathbf{J^P=\frac{1}{2}^\pm}$ baryonic fields, just like quark-antiquark mesonic fields, are parametrized
by $3\times 3$ matrices. In a chirally symmetric approach, it is also natural to
construct baryons from quarks and diquarks with definite behavior under
chiral transformations, i.e., from left- and right-handed quarks and diquarks.
If we want to include states that transform in the mirror assignment, such
that we can construct chirally invariant mass terms in the Lagrangian, we will
show that we are then necessarily lead to consider four distinct baryonic multiplets. 
The possibility to have four multiplets of chiral partners in the mirror assignment was 
already discussed in the outlook of Ref.\ \cite{Gallas:2009qp}. Then, instead
of only the ground-state baryon (the nucleon doublet for $N_{f}=2$) and its
chiral partner, two positive-parity baryons
(the nucleon and the Roper $N(1440)$ for $N_{f}=2$) and two
negative-parity states ($N(1535)$ and $N(1650)$ for $N_{f}=2$) occur.

This paper is organized as follows. In Sec.\ 2 we present our model and its
implications. Namely, in Sec.\ 2.1 we introduce the baryonic fields for $N_{f}=3$
and in Sec.\ 2.2 the corresponding Lagrangian.
A full $N_{f}=3$ analysis with $32=8\times4$ baryonic resonances is very difficult.
Therefore, for the present work we decided to study a simplified scenario by considering
a reduction of the $N_f=3$ Lagrangian to the $N_{f}=2$ case.
This reduction is discussed in Sec.\ 2.3. In Sec.\ 2.4 the mass matrix involving the
four nucleonic states $N(939),\, N(1440),\, N(1535),$ and $N(1650)$ is determined and
diagonalized. In Sec.\ 3 we perform a fit of the parameters of our model
to experimental data \cite{PDG} for the masses, decay widths, and axial coupling constants. 
In Sec.\ 4 we discuss our results and give an outlook to future work. Technical details are relegated to
various appendices.

We use natural units, $\hbar=c=1$, and the metric tensor is $(g^{\mu\nu})=\text{diag}(+,-,-,-)$.

\section{The Model and its Implications}

In this section we first construct the baryonic fields in a chiral quark-diquark picture.
We account for the fact that two of the four baryonic fields transform in a ``mirror'' way
as compared to the other two. We then present the complete Lagrangian of the eLSM 
for $N_{f}=3$ flavors. A reduction to $N_{f}=2$ flavors is performed and finally
the mass matrix for the four nucleonic states $N(939),\, N(1440),\, N(1535),$ and $N(1650)$
is given.

\subsection{Baryonic Fields for $N_{f}=3$}

\label{sec:Nf=3}

In the two-flavor case one works with isospin doublets $\psi_{i}$, where the
upper field is proton-like, i.e., of the type $uud$, and the lower field is
neutron-like, i.e., of the type $udd$. The right- and 
left-handed components $\psi_{iR}$ and $\psi_{iL}$ behave either
in a ``naive'' or in a mirror way under chiral transformations. 
The naive transformation behavior implies
$\psi_{iR}\rightarrow U_{R}\psi_{iR}$ and $\psi_{iL}\rightarrow U_{L}\psi
_{iL}$, while the mirror one implies $\psi_{iR}\rightarrow U_{L}\psi_{iR}$ and
$\psi_{iL}\rightarrow U_{R}\psi_{iL}$, where the index $i$ labels the
nucleonic doublets and the quantities $U_{R}$ and $U_{L}$ are $2\times2$ 
matrix representations of the chiral group $U(2)_{R}\times U(2)_{L}$.

As mentioned in the Introduction, for three flavors $J^P=\frac{1}{2}^+$ baryons
are described by $3\times3$ matrices. In order to construct these fields we use 
a chiral quark-diquark model [see Ref.\ \cite{quarkDiquark} and in particular
Ref.\ \cite{tqmix}], i.e., baryons are considered to be made of a quark and a diquark,
where a diquark is a (colored) state consisting of two quarks.
We are interested in so-called ``good'' diquarks \cite{jaffe,tetraquarks,tqmix}
which are (pseudo)scalar objects with antisymmetric color- and flavor-wave function.
For $N_{f}=3$ there are three scalar, $J^{P}=0^{+}$, and three pseudoscalar
diquarks, $J^{P}=0^{-}$. Mathematically they can be expressed as follows
\cite{tqmix}:
\begin{align}
J^{P}=0^{+}:\qquad\quad\mathcal{D}_{ij}  &  =\frac{1}{\sqrt{2}}\left(
q_{j}^{T}C\gamma^{5}q_{i}-q_{i}^{T}C\gamma^{5}q_{j}\right)  \equiv\sum
_{k=1}^{3}D_{k}\epsilon_{kij}\qquad\text{with}\quad D_{k}=\frac{1}{\sqrt{2}%
}\epsilon_{klm}q_{m}^{T}C\gamma^{5}q_{l}\;, \nonumber \\[-0.3cm]
J^{P}=0^{-}:\qquad\quad\tilde{\mathcal{D}}_{ij}  &  =\frac{1}{\sqrt{2}}\left(
q_{j}^{T}Cq_{i}-q_{i}^{T}Cq_{j}\right)  \equiv\sum_{k=1}^{3}\tilde{D}%
_{k}\epsilon_{kij}\qquad\qquad\text{with}\quad\tilde{D}_{k}=\frac{1}{\sqrt{2}%
}\epsilon_{klm}q_{m}^{T}Cq_{l}\;, \label{diquarks}
\end{align}
where $D_{k}$ is the scalar diquark current and $\tilde{D}_{k}$ is the
pseudoscalar diquark current. The indices $i,j,k,l$, and $m$ are flavor indices.
The color structure of these objects is formally identical to the flavor structure
and thus suppressed here. From the scalar and pseudoscalar diquarks 
(\ref{diquarks}) we can construct left- and right-handed diquarks,
\begin{align*}
\mathcal{D}_R &:= \frac{1}{\sqrt{2}}\left(\tilde{\mathcal{D}} + \mathcal{D}\right) = \sum_{i = 1}^{3}D_i^R A^{i} \qquad 
\text{with} \qquad D_i^R \equiv \frac{1}{\sqrt{2}}\left(\tilde{D}_i + D_i\right)\; ,\nonumber\\
\mathcal{D}_L &:= \frac{1}{\sqrt{2}}\left(\tilde{\mathcal{D}} - \mathcal{D}\right) = \sum_{i = 1}^{3}D_i^L A^{i} \qquad 
\text{with} \qquad D_i^L \equiv \frac{1}{\sqrt{2}}\left(\tilde{D}_i - D_i\right)\; ,
\end{align*}
where $(A_{i})_{jk}=\epsilon_{ijk}$. Under $U(3)_{L}\times U(3)_{R}$ chiral
transformations they behave as
\begin{align}
D_i^{L}\rightarrow D_i^{L}U_{L}^{\dagger}\;, \qquad
D_i^{R}\rightarrow D_i^{R}U_{R}^{\dagger}\;,
\label{eq:chiral trafo DIQUARK}
\end{align}
where $U_{L}$ and $U_{R}$ are unitary $3\times3$ matrices. Thus,
$D_{i}^{L(R)}$ transforms as a left-(right-)handed antiquark.

In order to construct baryonic fields as quark-diquark pairs, we have to combine $D_j^{R}$ or
$D_j^{L}$ with a quark, $q_i$,
\begin{align*}
N_{1}  &  \; \equiv \;(N_{1})_{ij}\quad\hat{=}\quad D^{R}_{j}q_{i}=\frac{1}{\sqrt{2}%
}\left(  \tilde{D}_{j}+D_{j}\right)  q_{i} \;,\\
N_{2}  &  \;\equiv \;(N_{2})_{ij}\quad\hat{=}\quad D^{L}_{j}q_{i}=\frac{1}{\sqrt{2}%
}\left(  \tilde{D}_{j}-D_{j}\right)  q_{i}\;.
\end{align*}
These two fields are obviously $3 \times3$ matrices in flavor space.

We now compute the left- and right-handed components of these fields. To this
end, one has to take into account that the chiral projection operators act only on the quark
fields $q_i$, because they carry a spinor index, while the diquarks are scalars in Dirac
space,
\[
N_{1(2)R}=\mathcal{P}_{R}N_{1(2)}\; \hat{=} \; D^{R(L)}q_{R}\; ,
\qquad N_{1(2)L}=\mathcal{P}_{L}N_{1(2)}\; \hat{=} \; D^{R(L)}q_{L}\;. 
\]
Using the transformation behavior of a quark spinor and Eq.\
(\ref{eq:chiral trafo DIQUARK}), the chiral transformation of the baryonic
fields can be computed as
\begin{align}
N_{1R}  &  \rightarrow~U_{R}N_{1R}U_{R}^{\dagger}\; ,\qquad N_{1L}%
\rightarrow~U_{L}N_{1L}U_{R}^{\dagger}\; ,\nonumber\\
N_{2R}  &  \rightarrow~U_{R}N_{2R}U_{L}^{\dagger}\; ,\qquad N_{2L}%
\rightarrow~U_{L}N_{2L}U_{L}^{\dagger}\; . \label{eq:chiralTrafo of N_1/2}%
\end{align}
One observes that the chiral transformation from the left follows the naive assignment, while the 
one from the right results from the transformation of the diquark field ($1 \leftrightarrow R$, $2\leftrightarrow L$).
Thus, the presence of two multiplets which transform in a naive way (from the left) 
is quite natural in the $N_{f}=3$ framework. 

The behavior under parity and charge-conjugation transformations is given by%
\begin{equation}%
\begin{tabular}
[c]{l|c|c}
& parity & charge conjugation\\\hline
&  & \\[-0.3cm]%
$N_{1R}~~~~~$ & $\quad-\gamma^{0}N_{2L}(t,-\boldsymbol{x})\quad$ &
$-i\gamma^{2}\left(  N_{2L}\right)  ^{\star}$\\
$N_{1L}$ & $\quad-\gamma^{0}N_{2R}(t,-\boldsymbol{x})\quad$ & $-i\gamma
^{2}\left(  N_{2R}\right)  ^{\star}$\\
$N_{2R}$ & $\quad-\gamma^{0}N_{1L}(t,-\boldsymbol{x})\quad$ & $-i\gamma
^{2}\left(  N_{1L}\right)  ^{\star}$\\
$N_{2L}$ & $\quad-\gamma^{0}N_{1R}(t,-\boldsymbol{x})\quad$ & $-i\gamma
^{2}\left(  N_{1R}\right)  ^{\star}$%
\end{tabular}
\end{equation}
which shows that the fields $N_{1}$ and $N_{2}$ are not parity eigenstates and
cannot be directly associated with existing resonances (even in the limit of vanishing mixing). 

Furthermore, we introduce two baryonic matrices $M_{1}$ and $M_{2}$ whose
chiral transformation from the left is ``mirror-like''. These fields can be constructed in the
same way as $N_{1}$ and $N_{2}$, however, we need to include an
additional Dirac matrix so that a left-(right-)handed projection operator is
converted into a right-(left-)handed one (due to the commutation relation 
$[\gamma^{5},\gamma^{\mu}]=0$). Only then one can act with a right-(left-)handed
chiral transformation $U_{R(L)}$ from the left onto $M_{i R(L)}$. To contract the additional Lorentz
index we also include a partial derivative. Consequently, the mathematical
structure of the ``mirror-like'' fields is given by
\begin{align*}
M_{1}  &  \; \equiv \; (M_{1})_{ij}\quad\hat{=}\quad D_{j}^{R}\gamma^{\mu}\partial_{\mu
}q_{i}=\frac{1}{\sqrt{2}}\left(  \tilde{D}_{j}+D_{j}\right)  \gamma^{\mu
}\partial_{\mu}q_{i}\; ,\\
M_{2}  &  \; \equiv \; (M_{2})_{ij}\quad\hat{=}\quad D_{j}^{L}\gamma^{\mu}\partial_{\mu
}q_{i}=\frac{1}{\sqrt{2}}\left(  \tilde{D}_{j}-D_{j}\right)  \gamma^{\mu
}\partial_{\mu}q_{i}\; .
\end{align*}
Their chiral transformations are given by
\begin{align}
M_{1R}  &  \rightarrow~U_{L}M_{1R}U_{R}^{\dagger}\; ,\qquad M_{1L}%
\rightarrow~U_{R}M_{1L}U_{R}^{\dagger} \; ,\nonumber\\
M_{2R}  &  \rightarrow~U_{L}M_{2R}U_{L}^{\dagger}\; ,\qquad M_{2L}%
\rightarrow~U_{R}M_{2L}U_{L}^{\dagger}\; , \label{eq:chiralTrafo of M_1/2}%
\end{align}
where the left-transformation is now mirror-like while the 
one from the right results from the transformation of the diquark field 
($1 \leftrightarrow R$, $2\leftrightarrow L$).
Under parity they transform just as $N_{1}$ and $N_{2}$, but under charge 
conjugation they transform with a reversed sign:
\begin{equation} \label{PCM}
\begin{tabular}
[c]{l|c|c}
& parity & charge conjugation\\\hline
&  & \\[-0.3cm]%
$M_{1R}~~~~~$ & $\quad-\gamma^{0}M_{2L}(t,-\boldsymbol{x})\quad$ &
$i\gamma^{2}\left(  M_{2L}\right)  ^{\star}$\\
$M_{1L}$ & $\quad-\gamma^{0}M_{2R}(t,-\boldsymbol{x})\quad$ & $i\gamma
^{2}\left(  M_{2R}\right)  ^{\star}$\\
$M_{2R}$ & $\quad-\gamma^{0}M_{1L}(t,-\boldsymbol{x})\quad$ & $i\gamma
^{2}\left(  M_{1L}\right)  ^{\star}$\\
$M_{2L}$ & $\quad-\gamma^{0}M_{1R}(t,-\boldsymbol{x})\quad$ & $i\gamma
^{2}\left(  M_{1R}\right)  ^{\star}$%
\end{tabular}
\end{equation}

The transformation laws (\ref{eq:chiralTrafo of N_1/2}) -- (\ref{PCM}) 
allow us to write down a baryonic Lagrangian with chirally
invariant mass terms, see next section and Appendix \ref{sec:Full Nf=3 Lag}.

Baryonic fields with definite behavior under parity transformations are introduced as:
\begin{equation}
B_{N}=\frac{N_{1}-N_{2}}{\sqrt{2}},\qquad B_{N\ast}=\frac{N_{1}+N_{2}}%
{\sqrt{2}},\qquad B_{M}=\frac{M_{1}-M_{2}}{\sqrt{2}},\qquad B_{M\ast}%
=\frac{M_{1}+M_{2}}{\sqrt{2}}\text{,} \label{eq:def of B fields}%
\end{equation}
where now $B_{N}$ and $B_{M}$ have positive parity and $B_{N\ast}$ and
$B_{M\ast}$ have negative parity. In the limit of zero mixing, $B_{N}$
describes the ground-state baryonic fields of Eq.\ (\ref{groundstatebaryons}),
i.e., $\{N(939)$, $\Lambda(1116),$ $\Sigma(1193),$ $\Xi(1338)\},$ $B_{M}$ the
positive-parity fields $\{N(1440),\Lambda(1600),\Sigma(1660),\Xi(1690)\}$,
$B_{N\ast}$ can be assigned to the negative-parity fields $\{N(1535),$
$\Lambda(1670),\Sigma(1620),\Xi(?)\}$ and, finally, $B_{M\ast}$ to 
$\{N(1650),\Lambda(1800),\Sigma(1750),\Xi(?)\}$. The detailed study of the mixing
will be performed below for the two-flavor case. 

\subsection{The eLSM Lagrangian for $N_{f}=3$}

The mesonic part of the Lagrangian of the eLSM reads \cite{denisnf3}:
\begin{align}
\mathcal{L}_{\text{meson}}=  &  \Tr\left\{  (D^{\mu}\Phi)^{\dagger}D_{\mu}%
\Phi)\right\}  -m_{0}^{2}\Tr\left\{  \Phi^{\dagger}\Phi\right\}  -\lambda
_{1}\left(  \Tr\left\{  \Phi^{\dagger}\Phi\right\}  \right)  ^{2}-\lambda
_{2}\Tr\left\{  \left(  \Phi^{\dagger}\Phi\right)  ^{2}\right\}  \nonumber\\
&  -\frac{1}{4}\Tr\left\{  L_{\mu\nu}L^{\mu\nu}+R_{\mu\nu}R^{\mu\nu}\right\}
+\Tr\left\{  \left(  \frac{m_{1}^2}{2}+\Delta\right)  \left(  L_{\mu}L^{\mu
}+R_{\mu}R^{\mu}\right)  \right\}  \nonumber\\
&  +\Tr\left\{  H\left(  \Phi+\Phi^{\dagger}\right)  \right\}  +c\left(
\det\Phi-\det\Phi^{\dagger}\right)  ^{2}\nonumber\\
&  +i\frac{g_{2}}{2}\left(  \Tr\left\{  L_{\mu\nu}\left[  L^{\mu},L^{\nu
}\right]  \right\}  +\Tr\left\{  R_{\mu\nu}\left[  R^{\mu},R^{\nu}\right]
\right\}  \right)  \nonumber\\
&  +\frac{h_{1}}{2}\Tr\left\{  \Phi^{\dagger}\Phi\right\}  \Tr\left\{  L_{\mu
}L^{\mu}+R_{\mu}R^{\mu}\right\}  +h_{2}\Tr\left\{  (L_{\mu}%
\Phi)^{\dagger}(L^{\mu}\Phi)+(\Phi R_{\mu})^{\dagger}(\Phi R^{\mu})\right\}
\nonumber\\
&  +2h_{3}\Tr\left\{  \Phi R^{\mu}\Phi^{\dagger}L^{\mu}\right\}  +g_{3}\left(
\Tr\left\{  L_{\mu}L_{\nu}L^{\mu}L^{\nu}\right\}  +\Tr\left\{  R_{\mu}R_{\nu
}R^{\mu}R^{\nu}\right\}  \right)  \nonumber\\
&  +g_{4}\left(  \Tr\left\{  L_{\mu}L^{\mu}L_{\nu}L^{\nu}\right\}
+\Tr\left\{  R_{\mu}R^{\mu}R_{\nu}R^{\nu}\right\}  \right)  +g_{5}\Tr\left\{
L_{\mu}L^{\mu}\right\}  \Tr\left\{  R_{\nu}R^{\nu}\right\}  \nonumber\\
&  +g_{6}\left(  \Tr\left\{  L_{\mu}L^{\mu}\right\}  \Tr\left\{  L_{\nu}%
L^{\nu}\right\}  +\Tr\left\{  R_{\mu}R^{\mu}\right\}  \Tr\left\{  R_{\nu
}R^{\nu}\right\}  \right)  \; , \label{eq:Lmeson}%
\end{align}
with the covariant derivative $D^{\mu}\Phi=\partial^{\mu}\Phi-ig_{1}\left(
L^{\mu}\Phi-\Phi R^{\mu}\right)  $ and the field-strength tensors $R^{\mu\nu
}=\partial^{\mu}R^{\nu}-\partial^{\nu}R^{\mu},~L^{\mu\nu}=\partial^{\mu}%
L^{\nu}-\partial^{\nu}L^{\mu}.$ The matrices $\Phi$, $R^{\mu}$, and $L^{\mu}$
represent the (pseudo)scalar and (axial-)vector nonets:
\begin{align}
\Phi &  =\sum_{i=0}^{8}(S_{i}+iP_{i})T_{i}=\frac{1}{\sqrt{2}}\left(
\begin{array}
[c]{ccc}%
\frac{\sigma_{N}+a_{0}^{0}+i(\eta_{N}+\pi^{0})}{\sqrt{2}} & a_{0}^{+}+i\pi^{+}
& K_{0}^{*+}+iK^{+}\\
a_{0}^{-}+i\pi^{-} & \frac{\sigma_{N}-a_{0}^{0}+i(\eta_{N}-\pi^{0})}{\sqrt{2}}
& K_{0}^{*0}+iK^{0}\\
K_{0}^{*-}+iK^{-} & \bar{K}_{0}^{*0}+i\bar{K}^{0} & \sigma_{S}+i\eta_{S}%
\end{array}
\right)  \; ,\nonumber\\
R^{\mu}  &  =\sum_{i=0}^{8}(V_{i}^{\mu}-A_{i}^{\mu})T_{i}=\frac{1}{\sqrt{2}%
}\left(
\begin{array}
[c]{ccc}%
\frac{\omega_{N}^{\mu}+\rho^{0\mu}}{\sqrt{2}}-\frac{f_{1N}^{\mu}+a_{1}^{0\mu}%
}{\sqrt{2}} & \rho^{+\mu}-a_{1}^{+\mu} & K^{\ast+\mu}-K_{1}^{+\mu}\\
\rho^{-\mu}-a_{1}^{-\mu} & \frac{\omega_{N}^{\mu}-\rho^{0\mu}}{\sqrt{2}}%
-\frac{f_{1N}^{\mu}-a_{1}^{0\mu}}{\sqrt{2}} & K^{0\ast\mu}-K_{1}^{0\mu}\\
K^{\ast-\mu}-K_{1}^{-\mu} & \bar{K}^{\ast0\mu}-\bar{K}_{1}^{0\mu} & \omega
_{S}^{\mu}-f_{1S}^{\mu}%
\end{array}
\right)  \;,\nonumber\\
L^{\mu}  &  =\sum_{i=0}^{8}(V_{i}^{\mu}+A_{i}^{\mu})T_{i}=\frac{1}{\sqrt{2}%
}\left(
\begin{array}
[c]{ccc}%
\frac{\omega_{N}^{\mu}+\rho^{0\mu}}{\sqrt{2}}+\frac{f_{1N}^{\mu}+a_{1}^{0\mu}%
}{\sqrt{2}} & \rho^{+\mu}+a_{1}^{+\mu} & K^{\ast+\mu}+K_{1}^{+\mu}\\
\rho^{-\mu}+a_{1}^{-\mu} & \frac{\omega_{N}^{\mu}-\rho^{0\mu}}{\sqrt{2}}%
+\frac{f_{1N}^{\mu}-a_{1}^{0\mu}}{\sqrt{2}} & K^{\ast0\mu}+K_{1}^{0\mu}\\
K^{\ast-\mu}+K_{1}^{-\mu} & \bar{K}^{\ast0\mu}+\bar{K}_{1}^{0\mu} & \omega
_{S}^{\mu}+f_{1S}^{\mu}%
\end{array}
\right)  \; .
\end{align}
Here, $S_{i}$ $(i=0,\dots,8)$ represents the scalar, $P_{i}$ the pseudoscalar,
$V_{i}$ the vector, and $A_{i}$ the axial-vector mesonic fields. The quantities
$T_{i}$ are the generators of $U(3)$. Under $U(N_{f})_{R}\times U(N_{f})_{L}$
chiral transformations $\Phi$ behaves as $\Phi\rightarrow U_{L}\Phi
U_{R}^{\dagger}$
and the left- and right-handed vector fields as $R^{\mu}\rightarrow
U_{R}R^{\mu}U_{R}^{\dagger}$ and $L^{\mu}\rightarrow U_{L}L^{\mu}%
U_{L}^{\dagger}$.

For $H=\Delta=c=0$ the Lagrangian $\mathcal{L}_{\text{meson}}$ is invariant
under global chiral $U(3)_{R}\times U(3)_{L}\left(  =U(3)_{V}\times U(3)_{A}\right)$ 
transformations. The $U(1)_{A}$ anomaly of QCD is parametrized by
$c\neq0$. The explicit breaking of $U(3)_{A}$ due to the nonzero quark masses
in the (pseudo)scalar and (axial-)vector sector is implemented by the terms
proportional to $H$ and $\Delta$, respectively. We assume isospin symmetry for
the $u$ and $d$ quarks to be exact. As a consequence, only the pure
nonstrange scalar-isoscalar field $\sigma_{N}$ and the pure strange
scalar-isoscalar field $\sigma_{S}$, carrying the same quantum numbers as the
vacuum, condense and have nonzero vacuum expectation values (VEVs), for more
details and for the values of all relevant parameters, see Ref.\ \cite{denis}.

To describe the baryonic degrees of freedom and their interactions with
mesons, we use the following Lagrangian which is invariant under
global chiral $U(3)_{R}\times U(3)_{L}$ as well as parity and charge-conjugation
transformations:
\begin{align}
\mathcal{L}_{N_{f}=3}=  &  \quad\Tr\left\{  \bar{N}_{1L}i\gamma_{\mu}%
D_{2L}^{\mu}N_{1L}+\bar{N}_{1R}i\gamma_{\mu}D_{1R}^{\mu}N_{1R}+\bar{N}%
_{2L}i\gamma_{\mu}D_{1L}^{\mu}N_{2L}+\bar{N}_{2R}i\gamma_{\mu}D_{2R}^{\mu
}N_{2R}\right\}  \nonumber\\
&  +\Tr\left\{  \bar{M}_{1L}i\gamma_{\mu}D_{4R}^{\mu}M_{1L}+\bar{M}%
_{1R}i\gamma_{\mu}D_{3L}^{\mu}M_{1R}+\bar{M}_{2L}i\gamma_{\mu}D_{3R}^{\mu
}M_{2L}+\bar{M}_{2R}i\gamma_{\mu}D_{4L}^{\mu}M_{2R}\right\}  \nonumber\\
&  -g_{N}\Tr\left\{  \bar{N}_{1L}\Phi N_{1R}+\bar{N}_{1R}\Phi^{\dagger}%
N_{1L}+\bar{N}_{2L}\Phi N_{2R}+\bar{N}_{2R}\Phi^{\dagger}N_{2L}\right\}
\nonumber\\
&  -g_{M}\Tr\left\{  \bar{M}_{1L}\Phi^{\dagger}M_{1R}+\bar{M}_{1R}\Phi
M_{1L}+\bar{M}_{2L}\Phi^{\dagger}M_{2R}+\bar{M}_{2R}\Phi M_{2L}\right\}
\nonumber\\
&  -m_{0,1}\Tr\bigl\{\bar{N}_{1L}M_{1R}+\bar{M}_{1R}N_{1L}+\bar{N}_{2R}%
M_{2L}+\bar{M}_{2L}N_{2R}\bigr\}\nonumber\\
&  -m_{0,2}\Tr\bigl\{\bar{N}_{1R}M_{1L}+\bar{M}_{1L}N_{1R}+\bar{N}_{2L}%
M_{2R}+\bar{M}_{2R}N_{2L}\bigr\} \nonumber\\
&  -\kappa_{1}\Tr\left\{  \bar{N}_{1R}\Phi^{\dagger}N_{2L}\Phi+\bar{N}%
_{2L}\Phi N_{1R}\Phi^{\dagger}\right\}  -\kappa_{1}^{\prime}\Tr\left\{
\bar{N}_{1L}\Phi N_{2R}\Phi+\bar{N}_{2R}\Phi^{\dagger}N_{1L}\Phi^{\dagger
}\right\}  \nonumber\\
&  -\kappa_{2}\Tr\left\{  \bar{M}_{1R}\Phi M_{2L}\Phi+\bar{M}_{2L}%
\Phi^{\dagger}M_{1R}\Phi^{\dagger}\right\}  -\kappa_{2}^{\prime}\Tr\left\{
\bar{M}_{1L}\Phi^{\dagger}M_{2R}\Phi+\bar{M}_{2R}\Phi M_{1L}\Phi^{\dagger
}\right\}  \nonumber\\
&  -\epsilon_{1}\left(  \Tr\left\{  \bar{N}_{1L}\Phi\right\}  \Tr\left\{
N_{2R}\Phi\right\}  +\Tr\left\{  \bar{N}_{2R}\Phi^{\dagger}\right\}
\Tr\left\{  N_{1L}\Phi^{\dagger}\right\}  \right)  \nonumber\\
&  -\epsilon_{2}\left(  \Tr\left\{  \bar{M}_{1R}\Phi\right\}  \Tr\left\{
M_{2L}\Phi\right\}  +\Tr\left\{  \bar{M}_{2L}\Phi^{\dagger}\right\}
\Tr\left\{  M_{1R}\Phi^{\dagger}\right\}  \right)  \nonumber\\
&  -\epsilon_{3}\Tr\bigl\{\Phi^{\dagger}\Phi\bigr\}\Tr\bigl\{\bar{N}%
_{1L}M_{1R}+\bar{M}_{1R}N_{1L}+\bar{N}_{2R}M_{2L}+\bar{M}_{2L}N_{2R}%
\bigr\}\nonumber\\
&  -\epsilon_{4}\Tr\bigl\{\Phi^{\dagger}\Phi\bigr\}\Tr\bigl\{\bar{N}%
_{1R}M_{1L}+\bar{M}_{1L}N_{1R}+\bar{N}_{2L}M_{2R}+\bar{M}_{2R}N_{2L}%
\bigr\} \;, \label{eq:LagNf=3}%
\end{align}
where the covariant derivatives are given by
\[
D_{kR}^{\mu}=\partial^{\mu}-ic_{k}R^{\mu}\; ,\qquad D_{kL}^{\mu}=\partial^{\mu
}-ic_{k}L^{\mu}\; , \qquad k=1, \ldots, 4\;,
\]
with dimensionless coupling constants $c_{1}, \ldots, c_{4}$, 
which determine the strength of baryon-baryon-(axial-)vector interactions. 
The interactions of the baryonic fields with (pseudo)scalar mesons are parametrized
by $g_{N}$ and $g_{M}$, which are also dimensionless. The terms proportional to
$\kappa_{1}, \kappa_{2}, \kappa_{1}^{\prime}, \kappa_{2}^{\prime}$ (and
$\epsilon_{i}$) are included because otherwise the baryonic fields become 
pairwise degenerate in mass (see Appendix \ref{sec:masses without kappas}).
Terms parametrized by $\epsilon_{i}$ are proportional to a product of two
traces. Such terms are large-$N_{c}$ suppressed (OZI rule) and will be
neglected in the following discussion. The explicit form of the Lagrangian in
terms of the parity eigenstates $B_{N}$, $B_{N\ast}$, $B_{M}$, and $B_{M\ast}$
is given in Appendix \ref{sec:Full Nf=3 Lag}.

Note that the terms in the first four lines of Eq.\ (\ref{eq:LagNf=3}) 
have naive scaling dimension four, and are thus dilatation-invariant.
The terms in the fifth and sixth lines have naive scaling dimension three. Thus, they
formally break dilatation symmetry, but
can be made dilation-invariant assuming that $m_{0,1}$ and $m_{0,2}$ are
proportional to a gluon and/or a four-quark condensate (with a dimensionless
proportionality constant). Such terms arise
from the (dilatation-invariant) interaction of a glueball and/or a four-quark state with baryons,
assuming that a spontaneous or explicit symmetry breaking mechanism induces
a non-vanishing VEV for the gluon and/or the four-quark field.  
For a more detailed description of how
one can render the mass term dilation-invariant by including a tetraquark see
e.g.\ Ref. \cite{tetraquarks}. 

The terms in the seventh to twelfth line of Eq.\ (\ref{eq:LagNf=3}) have naive
scaling dimension five and therefore also break dilatation symmetry. However, 
in this case the coupling constants $\kappa_i$, $\kappa_i'$, and $\epsilon_i$ 
would need to be proportional to inverse powers of a gluon and/or a 
four-quark condensate. Such terms can only arise from non-analytic interaction terms
between baryons and glueballs/four-quark states, which should be avoided
in a Lagrangian prescription. Nevertheless, these terms may also be considered as
effective four-point interactions arising from two (dilatation-invariant) three-point interaction 
vertices between a meson, a baryon, and a heavier baryonic resonance, 
where the vertices are connected by a propagator of the latter. If the mass of the baryon resonance is
much larger than the typical energy scale where the Lagrangian (\ref{eq:LagNf=3}) is applicable,
its propagator may be considered to be static and homogeneous, resulting in the
four-point interactions proportional to $\kappa_i$, $\kappa_i'$, and $\epsilon_i$ 
in Eq.\ (\ref{eq:LagNf=3}).

\subsection{The Lagrangian for $N_{f}=2$}

In this section we reduce the $N_{f}=3$ Lagrangian (\ref{eq:LagNf=3})
to $N_{f}=2$ flavors. In order to achieve this reduction, we set all strange quark fields $s$
to zero. Only the (1 3)- and (2 3)-elements of the baryonic matrices
remain:
\begin{align}
B_{N}~~  &  \xrightarrow{ s = 0}\left(
\begin{array}
[c]{ccc}%
0 & 0 & \Psi_{N}^{1}\\
0 & 0 & \Psi_{N}^{2}\\
0 & 0 & 0
\end{array}
\right)  ,\qquad\qquad B_{N\ast}\quad\xrightarrow{ s = 0}\left(
\begin{array}
[c]{ccc}%
0 & 0 & \Psi_{N\ast}^{1}\\
0 & 0 & \Psi_{N\ast}^{2}\\
0 & 0 & 0
\end{array}
\right)  ,\\
B_{M}  &  \xrightarrow{ s = 0}\left(
\begin{array}
[c]{ccc}%
0 & 0 & \Psi_{M}^{1}\\
0 & 0 & \Psi_{M}^{2}\\
0 & 0 & 0
\end{array}
\right)  ,\qquad\qquad B_{M\ast}\xrightarrow{ s = 0}\left(
\begin{array}
[c]{ccc}%
0 & 0 & \Psi_{M\ast}^{1}\\
0 & 0 & \Psi_{M\ast}^{2}\\
0 & 0 & 0
\end{array}
\right)  ,
\end{align}
where $\Psi_{i}^{1(2)}$ $(i=N,N\ast,M,M\ast)$ are fields with quark content
$\Psi_{i}^{1}~\hat{=}~uud$ and $\Psi_{i}^{2}~\hat{=}~udd$. Applying the same
to the meson matrix $\Phi$ and to the left- and right-handed (axial-)vector
fields, $L^{\mu}$ and $R^{\mu}$, we obtain
\begin{align}
\Phi &  \xrightarrow{ S = 0 }\frac{1}{\sqrt{2}}\left(
\begin{array}
[c]{ccc}%
\frac{(\sigma_{N}+\varphi_{N}+a_{0}^{0})+i(\eta_{N}+\pi^{0})}{\sqrt{2}} &
a_{0}^{+}+i\pi^{+} & 0\\
a_{0}^{-}+i\pi^{-} & \frac{(\sigma_{N}+\varphi_{N}-a_{0}^{0})+i(\eta_{N}%
-\pi^{0})}{\sqrt{2}} & 0\\
0 & 0 & \varphi_{S}%
\end{array}
\right)  \equiv\left(
\begin{array}
[c]{cc}%
\Bigl(\Phi_{N_{f}=2}\Bigr) &
\begin{array}
[c]{c}%
0\\
0
\end{array}
\\
0\qquad0 & \frac{1}{\sqrt{2}}\varphi_{S}%
\end{array}
\right)  ,\\
R^{\mu}  &  \xrightarrow{ S = 0 }\frac{1}{\sqrt{2}}\left(
\begin{array}
[c]{ccc}%
\frac{\omega_{N}^{\mu}+\rho^{\mu0}}{\sqrt{2}}-\frac{f_{1N}^{\mu}+a_{1}^{\mu0}%
}{\sqrt{2}} & \rho^{\mu+}-a_{1}^{\mu+} & 0\\
\rho^{\mu-}-a_{1}^{\mu-} & \frac{\omega_{N}^{\mu}-\rho^{\mu0}}{\sqrt{2}}%
-\frac{f_{1N}^{\mu}-a_{1}^{\mu0}}{\sqrt{2}} & 0\\
0 & 0 & 0
\end{array}
\right)  \equiv\left(
\begin{array}
[c]{cc}%
\Bigl(R_{N_{f}=2}^{\mu}\Bigr) &
\begin{array}
[c]{c}%
0\\
0
\end{array}
\\
0\qquad0 & 0
\end{array}
\right)  ,\\
L^{\mu}  &  \xrightarrow{ S = 0 }\frac{1}{\sqrt{2}}\left(
\begin{array}
[c]{ccc}%
\frac{\omega_{N}^{\mu}+\rho^{\mu0}}{\sqrt{2}}+\frac{f_{1N}^{\mu}+a_{1}^{\mu0}%
}{\sqrt{2}} & \rho^{\mu+}+a_{1}^{\mu+} & 0\\
\rho^{\mu-}+a_{1}^{\mu-} & \frac{\omega_{N}^{\mu}-\rho^{\mu0}}{\sqrt{2}}%
+\frac{f_{1N}^{\mu}-a_{1}^{\mu0}}{\sqrt{2}} & 0\\
0 & 0 & 0
\end{array}
\right)  \equiv\left(
\begin{array}
[c]{cc}%
\Bigl(L_{N_{f}=2}^{\mu}\Bigr) &
\begin{array}
[c]{c}%
0\\
0
\end{array}
\\
0\qquad0 & 0
\end{array}
\right)  \text{.}%
\end{align}
Note that it is crucial to first consider the condensation of both scalar
fields $\sigma_{N}$ and $\sigma_{S}$ and only then set the mesons with $s$
quarks to zero, otherwise one would lose the VEV
$\varphi_{S}$ of the field $\sigma_{S}$. For $N_{f}=2$ it is common to write the
$2\times2$ meson matrices in the basis of the three $SU(2)$ generators
$\boldsymbol{T}=\boldsymbol{\tau}/2$, where $\boldsymbol{\tau}$ are the Pauli
matrices, and $T^{0}=\mathbbm{1}_{2x2}/2$:
\begin{align*}
\Phi_{N_{f}=2}  &  =\left(  \sigma_{N}+\varphi_{N}+i\eta_{N}\right)
T^{0}+\left(  \boldsymbol{a}_{0}+i\boldsymbol{\pi}\right)  \cdot
\boldsymbol{T}\;,\\
R_{N_{f}=2}^{\mu}  &  =\left(  \omega^{\mu}-f_{1}^{\mu}\right)  T^{0}+\left(
\boldsymbol{\rho}^{\mu}-\boldsymbol{a}_{1}^{\mu}\right)  \cdot\boldsymbol{T}\;,\\
L_{N_{f}=2}^{\mu}  &  =\left(  \omega^{\mu}+f_{1}^{\mu}\right)  T^{0}+\left(
\boldsymbol{\rho}^{\mu}+\boldsymbol{a}_{1}^{\mu}\right)  \cdot\boldsymbol{T}\;.
\end{align*}
As already indicated in the notation the fields are identified with the mesons
listed in Ref.\ \cite{PDG} in the following way. The scalar resonances $\sigma$
and $\boldsymbol{a}_{0}$ are assigned to $f_{0}(1370)$ and $a_{0}(1450)$.
The second possibility $\{\sigma,\boldsymbol{a}_{0}\}\hat{=}%
\{f_{0}(500),a_{0}(980)\}$ has to be excluded, because then our model cannot describe
the scattering lengths and the decay $\sigma\rightarrow\pi\pi$ at the same
time, for more details see Ref.\ \cite{denis}. The pseudoscalar $\eta_{N}\equiv(\bar{u}%
u+\bar{d}d)/\sqrt{2}$ is the $SU(2)$ counterpart of the $\eta$ meson, and
$\boldsymbol{\pi}$ corresponds to the pion triplet. The vectors $\omega^{\mu}$
and $\boldsymbol{\rho}^{\mu}$ represent the resonances $\omega(782)$ and
$\rho(770)$ and the axial-vector fields $f_{1}^{\mu}$ and $\boldsymbol{a}_{1}%
^{\mu}$ are identified with the resonances $f_{1}(1285)$ and $a_{1}(1260)$.

The resulting Lagrangian for the case $N_{f}=2$ reads (for details see Appendix
\ref{sec:Full Nf=3 Lag}):
\begin{align}
\mathcal{L}_{N_{f}=2}=~~  &  \bar{\Psi}_{NR}i\gamma_{\mu}D_{NR}^{\mu}\Psi
_{NR}+\bar{\Psi}_{NL}i\gamma_{\mu}D_{NL}^{\mu}\Psi_{NL}+\bar{\Psi}_{N\ast
R}i\gamma_{\mu}D_{NR}^{\mu}\Psi_{N\ast R}+\bar{\Psi}_{N\ast L}i\gamma_{\mu
}D_{NL}^{\mu}\Psi_{N\ast L}\nonumber\\
+  &  \bar{\Psi}_{MR}i\gamma_{\mu}D_{ML}^{\mu}\Psi_{MR}+\bar{\Psi}_{ML}%
i\gamma_{\mu}D_{MR}^{\mu}\Psi_{ML}+\bar{\Psi}_{M\ast R}i\gamma_{\mu}%
D_{ML}^{\mu}\Psi_{M\ast R}+\bar{\Psi}_{M\ast L}i\gamma_{\mu}D_{MR}^{\mu}%
\Psi_{M\ast L}\nonumber\\
+  &  c_{A_{N}}\bigl(\bar{\Psi}_{NR}i\gamma_{\mu}R^{\mu}\Psi_{N\ast R}%
+\bar{\Psi}_{N\ast R}i\gamma_{\mu}R^{\mu}\Psi_{NR}-\bar{\Psi}_{NL}i\gamma
_{\mu}L^{\mu}\Psi_{N\ast L}-\bar{\Psi}_{N\ast L}i\gamma_{\mu}L^{\mu}\Psi
_{NL}\bigr)\nonumber\\
+  &  c_{A_{M}}\bigl(\bar{\Psi}_{MR}i\gamma_{\mu}L^{\mu}\Psi_{M\ast R}%
+\bar{\Psi}_{M\ast R}i\gamma_{\mu}L^{\mu}\Psi_{MR}-\bar{\Psi}_{ML}i\gamma
_{\mu}R^{\mu}\Psi_{M\ast L}-\bar{\Psi}_{M\ast L}i\gamma_{\mu}R^{\mu}\Psi
_{ML}\bigr)\nonumber\\
-  &  g_{N}\bigl(\bar{\Psi}_{NL}\Phi\Psi_{NR}+\bar{\Psi}_{NR}\Phi^{\dagger
}\Psi_{NL}+\bar{\Psi}_{N\ast L}\Phi\Psi_{N\ast R}+\bar{\Psi}_{N\ast L}%
\Phi^{\dagger}\Psi_{N\ast R}\bigr)\nonumber\\
-  &  g_{M}\bigl(\bar{\Psi}_{ML}\Phi^{\dagger}\Psi_{MR}+\bar{\Psi}_{MR}%
\Phi\Psi_{ML}+\bar{\Psi}_{M\ast L}\Phi^{\dagger}\Psi_{M\ast R}+\bar{\Psi
}_{M\ast L}\Phi\Psi_{M\ast R}\bigr)\nonumber\\
-  &  \frac{m_{0,1}+m_{0,2}}{2}~\bigl(\bar{\Psi}_{NL}\Psi_{MR}+\bar{\Psi}%
_{NR}\Psi_{ML}+\bar{\Psi}_{N\ast L}\Psi_{M\ast R}+\bar{\Psi}_{N\ast R}%
\Psi_{M\ast L}\nonumber\\[-0.2cm]
&  \qquad\qquad\qquad~+\bar{\Psi}_{ML}\Psi_{NR}+\bar{\Psi}_{MR}\Psi_{NL}%
+\bar{\Psi}_{M\ast L}\Psi_{N\ast R}+\bar{\Psi}_{M\ast R}\Psi_{N\ast
L}\bigr)\nonumber\\[-0.2cm]
-  &  \frac{m_{0,1}-m_{0,2}}{2}~\bigl(\bar{\Psi}_{NL}\Psi_{M\ast R}-\bar{\Psi
}_{NR}\Psi_{M\ast L}-\bar{\Psi}_{ML}\Psi_{N\ast R}+\bar{\Psi}_{MR}\Psi_{N\ast
L}\nonumber\\[-0.2cm]
&  \qquad\qquad\qquad~-\bar{\Psi}_{N\ast L}\Psi_{MR}+\bar{\Psi}_{N\ast R}%
\Psi_{ML}+\bar{\Psi}_{M\ast L}\Psi_{NR}-\bar{\Psi}_{M\ast R}\Psi
_{NL}\bigr)\nonumber\\
-  &  \frac{\kappa_{1}^{\prime}+\kappa_{1}}{2}\frac{\varphi_{S}}{\sqrt{2}%
}~\bigl(-\bar{\Psi}_{NL}\Phi\Psi_{NR}-\bar{\Psi}_{NR}\Phi^{\dagger}\Psi
_{NL}+\bar{\Psi}_{N\ast L}\Phi\Psi_{N\ast R}+\bar{\Psi}_{N\ast R}\Phi
^{\dagger}\Psi_{N\ast L}\bigr)\nonumber\\[-0.15cm]
-  &  \frac{\kappa_{1}^{\prime}-\kappa_{1}}{2}\frac{\varphi_{S}}{\sqrt{2}%
}~\bigl(\bar{\Psi}_{NL}\Phi\Psi_{N\ast R}-\bar{\Psi}_{NR}\Phi^{\dagger}%
\Psi_{N\ast L}-\bar{\Psi}_{N\ast L}\Phi\Psi_{NR}+\bar{\Psi}_{N\ast R}%
\Phi^{\dagger}\Psi_{NL}\bigr)\nonumber\\[-0.15cm]
-  &  \frac{\kappa_{2}^{\prime}+\kappa_{2}}{2}\frac{\varphi_{S}}{\sqrt{2}%
}~\bigl(-\bar{\Psi}_{ML}\Phi^{\dagger}\Psi_{MR}-\bar{\Psi}_{MR}\Phi\Psi
_{ML}+\bar{\Psi}_{M\ast L}\Phi^{\dagger}\Psi_{M\ast R}+\bar{\Psi}_{M\ast
R}\Phi\Psi_{M\ast L}\bigr)\nonumber\\[-0.15cm]
-  &  \frac{\kappa_{2}^{\prime}-\kappa_{2}}{2}\frac{\varphi_{S}}{\sqrt{2}%
}~\bigl(\bar{\Psi}_{ML}\Phi^{\dagger}\Psi_{M\ast R}-\bar{\Psi}_{MR}\Phi
\Psi_{M\ast L}-\bar{\Psi}_{M\ast L}\Phi^{\dagger}\Psi_{MR}+\bar{\Psi}_{M\ast
R}\Phi\Psi_{ML}\bigr) \; ,\label{eq:LagNf=2}%
\end{align}
where we suppressed the subscript \textquotedblleft$N_{f}=2$\textquotedblright%
\ of the mesonic fields and introduced the isovectors $\Psi_{k}=(\Psi_{k}%
^{1},\Psi_{k}^{2})^{T}$, $k=N,N\ast,M,M\ast$. The covariant derivatives are
\[
D_{NR}^{\mu}=\partial^{\mu}-ic_{N}R^{\mu}~\text{,}\qquad D_{NL}^{\mu}%
=\partial^{\mu}-ic_{N}L^{\mu}\;, %
\]%
\[
D_{MR}^{\mu}=\partial^{\mu}-ic_{M}R^{\mu}~\text{,}\qquad D_{ML}^{\mu}%
=\partial^{\mu}-ic_{M}L^{\mu}\;, %
\]
with
\[
c_{N}=\frac{c_{1}+c_{2}}{2}\quad\text{ and }\quad c_{M}=\frac{c_{3}+c_{4}}%
{2}\; .
\]
These two constants parametrize the coupling between baryons of equal parity.
The constants
\[
c_{A_{N}}=\frac{c_{1}-c_{2}}{2}\quad\text{ and }\quad c_{A_{M}}=\frac
{c_{3}-c_{4}}{2}%
\]
describe the coupling of two baryons with different parity to (axial-)vector mesons.

Interestingly, the number of parameters of this $N_{f}=2$ Lagrangian  
obtained as a reduction of the more general $N_{f}=3$ Lagrangian is smaller
than what one would obtain by directly writing down the corresponding
two-flavor Lagrangian with four multiplets. This is due to the fact that some
terms are not allowed because of the more complex parity and charge-conjugation transformations 
of the baryonic fields in the $N_f = 3$ case [some terms which in principle have different 
coupling constants in the $N_{f}=2$ case \cite{Gallas:2009qp},
have now the same, as they transform into each other under parity or charge conjugation].

\subsection{The Mass Matrix}

\label{sec:mass matrix}

After SSB in the meson sector (see Appendix \ref{sec:Full Nf=2 Lag}), the
following terms contribute to the mass matrix of the four fields $\Psi_{N}$, $\Psi_{N\ast}$,
$\Psi_{M}$, and $\Psi_{M\ast}$:
\begin{align}
\mathcal{L}_{\text{mass}}  &  = -\left(  \frac{g_{N}\varphi_{N}}{2}
-\frac{\kappa_{1}^{\prime} + \kappa_{1}}{4\sqrt{2}}\varphi_{N}\varphi
_{S}\right)  \bar{\Psi}_{N}\Psi_{N} -\left(  \frac{g_{N}\varphi_{N}}{2}%
+\frac{\kappa_{1}^{\prime} + \kappa_{1}}{4\sqrt{2}}\varphi_{N}\varphi
_{S}\right)  \bar{\Psi}_{N\ast}\Psi_{N\ast}\nonumber\\
&  \quad-\left(  \frac{g_{M}\varphi_{N}}{2}-\frac{\kappa_{2}^{\prime} +
\kappa_{2}}{4\sqrt{2}}\varphi_{N}\varphi_{S}\right)  \bar{\Psi}_{M}\Psi_{M}
-\left(  \frac{g_{M}\varphi_{N}}{2}+\frac{\kappa_{2}^{\prime} + \kappa_{2}%
}{4\sqrt{2}}\varphi_{N}\varphi_{S}\right)  \bar{\Psi}_{M\ast}\Psi_{M\ast
}\nonumber\\
&  \quad- \frac{\kappa_{1}^{\prime} - \kappa_{1}}{4\sqrt{2}}\varphi_{N}%
\varphi_{S} \bigl( \bar{\Psi}_{N}\gamma^{5}\Psi_{N\ast} - \bar{\Psi}_{N\ast
}\gamma^{5}\Psi_{N} \bigr) - \frac{\kappa_{2}^{\prime} - \kappa_{2}}{4\sqrt
{2}}\varphi_{N}\varphi_{S} \bigl( \bar{\Psi}_{M}\gamma^{5}\Psi_{M\ast} -
\bar{\Psi}_{M\ast}\gamma^{5}\Psi_{M} \bigr)\nonumber\\
&  \quad-\frac{m_{0,1}+m_{0,2}}{2}\Bigl( \bar{\Psi}_{N}\Psi_{M}+ \bar{\Psi
}_{N\ast}\Psi_{M\ast}+ \bar{\Psi}_{M}\Psi_{N}+ \bar{\Psi}_{M\ast}\Psi_{N\ast
}\Bigr)\nonumber\\
&  \quad-\frac{m_{0,2}-m_{0,1}}{2}\Bigl( \bar{\Psi}_{N}\gamma^{5}\Psi_{M\ast}+
\bar{\Psi}_{N\ast}\gamma^{5}\Psi_{M}- \bar{\Psi}_{M}\gamma^{5}\Psi_{N\ast}-
\bar{\Psi}_{M\ast}\gamma^{5}\Psi_{N} \Bigr)\;,
\label{eq:mass Lagrangian}%
\end{align}
where $\varphi_{N}$ and $\varphi_{S}$ are the VEVs 
of the $\sigma_{N}$ and $\sigma_{S}$ meson, respectively. In order to
determine the physical fields $N_{939}$, $N_{1535}$, $N_{1440}$, and
$N_{1650}$ corresponding to the resonances $N(939),$ $N(1525),$ $N(1535),$
and $N(1640),$ we have to diagonalize the Lagrangian. To this end, we define
the vector
\begin{equation}
\Psi=(\Psi_{N},\gamma^{5}\Psi_{N\ast},\Psi_{M},\gamma^{5}\Psi_{M\ast}%
)^{T}\qquad\qquad\Longrightarrow\qquad\qquad\bar{\Psi}=(\bar{\Psi}_{N}%
,-\bar{\Psi}_{N\ast}\gamma^{5},\bar{\Psi}_{M},-\bar{\Psi}_{M\ast}\gamma^{5})\; .
\label{eq:Psi}%
\end{equation}
The additional $\gamma^{5}$ matrices are introduced in order to avoid such
matrices in the mass matrix (\ref{eq:mass matrix}).
As a consequence, all four components of the vector $\Psi$ have
the same parity. 

Rewriting Eq.\ (\ref{eq:mass Lagrangian}) in matrix form,
$\mathcal{L}_{\text{mass}}  =-\bar{\Psi}M\Psi$,
we obtain the mass matrix
\begin{align}
M \equiv \frac{1}{2}\left(
\begin{array}
[c]{cccc}%
g_{N}\varphi_{N}-\frac{\kappa_{1}^{\prime} + \kappa_{1}}{2\sqrt{2}}\varphi
_{N}\varphi_{S} & \frac{\kappa_{1}^{\prime} - \kappa_{1}}{2\sqrt{2}}%
\varphi_{N}\varphi_{S} & m_{0,1}+m_{0,2} & m_{0,1}-m_{0,2}\\
\frac{\kappa_{1}^{\prime} - \kappa_{1}}{2\sqrt{2}}\varphi_{N}\varphi_{S} &
-g_{N}\varphi_{N}-\frac{\kappa_{1}^{\prime} + \kappa_{1}}{2\sqrt{2}}%
\varphi_{N}\varphi_{S} & m_{0,2}-m_{0,1} & -m_{0,1}-m_{0,2}\\
m_{0,1}+m_{0,2} & m_{0,2}-m_{0,1} & g_{M}\varphi_{N}-\frac{\kappa_{2}^{\prime}
+ \kappa_{2}}{2\sqrt{2}}\varphi_{N}\varphi_{S} & \frac{\kappa_{2}^{\prime} -
\kappa_{2}}{2\sqrt{2}}\varphi_{N}\varphi_{S}\\
m_{0,1}-m_{0,2} & -m_{0,1}-m_{0,2} & \frac{\kappa_{2}^{\prime} - \kappa_{2}%
}{2\sqrt{2}}\varphi_{N}\varphi_{S} & -g_{M}\varphi_{N}-\frac{\kappa
_{2}^{\prime} + \kappa_{2}}{2\sqrt{2}}\varphi_{N}\varphi_{S}%
\end{array}
\right)\;.  \label{eq:mass matrix}%
\end{align}


At this point it is possible to compare to the Lagrangian of Ref.\
\cite{Gallas:2009qp}, which only describes the nucleon and its chiral partner.
If $m_{0,1}=-m_{0,2}$ and $\kappa_{1(2)}=\kappa_{1(2)}^{\prime}%
$, the mass matrix is of the form,
\[
M_{\text{decoupled}}= \frac{1}{2}\left(
\begin{array}
[c]{cccc}%
g_{N}\varphi_{N}-\frac{\kappa_{1}}{\sqrt{2}}\varphi
_{N}\varphi_{S} & 0 & 0 & 2m_{0,1}\\
0 & -g_{N}\varphi_{N}-\frac{\kappa_{1}}{\sqrt{2}}%
\varphi_{N}\varphi_{S} & -2m_{0,1} & 0\\
0 & -2m_{0,1} & g_{M}\varphi_{N}-\frac{\kappa_{2}}{\sqrt{2}}\varphi_{N}\varphi_{S} & 0\\
2m_{0,1} & 0 & 0 & -g_{M}\varphi_{N}-\frac{\kappa
_{2}}{\sqrt{2}}\varphi_{N}\varphi_{S}%
\end{array}
\right)\;. 
\]
Obviously, the fields $\Psi_{N}$ and $\Psi_{M\ast}$ completely decouple from the fields
$\Psi_{N\ast}$ and $\Psi_{M}$, and the diagonalization of the two sets can be
performed independently. However, it is not clear which of the two
states $\Psi_{N\ast}$ and $\Psi_{M\ast}$ should be identified with the chiral partner
of $\Psi_{N}$ (the putative nucleon field), because all states become degenerate in 
mass (all masses are equal to $m_{0,1}$) when chiral symmetry is restored 
($\varphi_N, \varphi_S \rightarrow 0$).

In order to diagonalize the mass matrix
(\ref{eq:mass matrix}), we have to solve the eigenvalue problem
\begin{align}
M\boldsymbol{u}_{k}  &  =m_{k}\boldsymbol{u}_{k}\; ,\nonumber\\
M^{ij}u_{k}^{j}  &  =m_{k}u_{k}^{i}\; , \label{eq:EVproblem}%
\end{align}
where $\boldsymbol{u}_{k}$ ($k\in\{1,...,\text{dim}(M)=4\}$) are the
eigenvectors and $m_{k}$ are the four eigenvalues of the mass matrix $M$.
Note that a sum over $j$ (but not over $k$) is understood. By multiplying
Eq.\ (\ref{eq:EVproblem}) with $\boldsymbol{u}_{l}$ from the left-hand side, we
find
\[
u_{l}^{i}M_{ij}u_{k}^{j}=m_{k}u_{l}^{i}u_{k}^{i} \equiv m_{k}\delta_{kl}\; ,
\]
for orthogonal eigenvectors, $\boldsymbol{u}_{l}\cdot\boldsymbol{u}_{k}=\delta_{lk}$. 
Hence the matrix
\begin{equation}
U_{ij}=u_{j}^{i} \label{eq:U}%
\end{equation}
diagonalizes $M$:
\[
U^{\dagger}MU=\text{diag}(m_{1},m_{2},m_{3},m_{4})~\equiv~\text{diag}%
(m_{939},-m_{1535},m_{1440},-m_{1650})\; .
\]
In the second equality we took into account that, due to the
definitions (\ref{eq:Psi}) and (\ref{eq:Psi phys}), the masses of the
negative-parity states correspond to negative eigenvalues of $M$.
Returning to the Lagrangian
 (\ref{eq:mass Lagrangian}), we now realize
that it is diagonalized by
\[
\mathcal{L}_{\text{mass}}=-\bar{\Psi}UU^{\dagger}MUU^{\dagger}\Psi=-\bar{\Psi
}^{\text{phys}}\text{diag}(m_{1},m_{2},m_{3},m_{4})\Psi^{\text{phys}}\; ,
\]
with the physical fields
\begin{equation}
\Psi^{\text{phys}}=U^{\dagger}\Psi~~\equiv~\left(  N_{939},\gamma^{5}%
N_{1535},N_{1440},\gamma^{5}N_{1650}\right)  ^{T}\; . \label{eq:Psi phys}%
\end{equation}
The eigenvalues of $M$, which (up to a sign) correspond to the masses of
the physical fields $N_{939},N_{1535},N_{1440}$, and $N_{1650}$, are determined
by the roots of the equation
\begin{align*}
\det\Bigl[M-m_{i}\mathbbm{1}_{4\times
4}\Bigr]{=}0\; .
\end{align*}
In the general case, this will be done numerically, see Sec.\ \ref{sec:fit}. However, 
in the chiral limit, i.e., $\varphi_N, \varphi_S \rightarrow 0$, one can easily do this analytically. 
In this case, denoting $\bar{M} \equiv (m_{0,1} + m_{0,2})/2$ and $\mu \equiv (m_{0,1} - m_{0,2})/2$,
the mass matrix reads
\[
M_{\text{chiral limit}} \equiv\left(
\begin{array}
[c]{cccc}%
0 &  0  & \bar{M} & \mu\\
0 & 0  & -\mu & -\bar{M}\\
\bar{M} & -\mu & 0 & 0\\
\mu & -\bar{M} & 0 & 0
\end{array}
\right)\;. 
\]
The eigenvalues of this matrix are $\lambda_{1,2} = \pm (\bar{M}+\mu) = \pm m_{0,1}$ 
and $\lambda_{3,4}= \pm (\bar{M}-\mu)= \pm m_{0,2}$. As expected, we have two distinct sets
of chiral partners. One set has the mass $m_{0,1}$ and the other the mass $m_{0,2}$, which
is in general different from $m_{0,1}$. In order to decide which mass eigenstates are
chiral partners, we need to compute the transformation matrix $U$. Somewhat surprisingly,
\[
U = \frac{1}{2} \left( \begin{array}{rrrr}
1 & -1 & 1 & 1 \\
-1 & 1 &1 & 1 \\
1 & 1 & 1 & -1 \\
1 & 1 & -1 & 1
\end{array} \right) \equiv U^\dagger \;,
\]
which means that the mass eigenstates are uniform
mixtures of the fields $\Psi_N, \gamma_5 \Psi_{N*}, \Psi_M,$ and $\gamma_5 \Psi_{M*}$.
The chiral partners with mass $m_{0,1}$ are given by the linear combinations
$\Psi_N - \gamma_5\Psi_{N*} + (\Psi_M + \gamma_5 \Psi_{M*})$ and
$-\Psi_N + \gamma_5\Psi_{N*} + (\Psi_M + \gamma_5 \Psi_{M*})$, while the
chiral partners with mass $m_{0,2}$ are given by
$\Psi_N + \gamma_5\Psi_{N*} + (\Psi_M - \gamma_5 \Psi_{M*})$ and
$\Psi_N + \gamma_5\Psi_{N*} - (\Psi_M - \gamma_5 \Psi_{M*})$, respectively.
Therefore, it is impossible to decide whether $N(1535)$ or $N(1650)$ is the
chiral partner of the nucleon. The solution to this problem will be presented in the next section, where we
compute the eigenvalues as a function of $\varphi_N$ to trace whether the mass of
$N(1535)$ or that of $N(1650)$ approaches the nucleon mass in the chiral limit.


\section{Results}
\label{sec:fit} 

The Lagrangian of the model in the $N_{f}=2$ case
[cf.\ Eqs.\ (\ref{eq:LagNf=2}) and (\ref{eq:full Lag Nf=2})] contains the following
twelve parameters in the baryonic sector: the mass parameters
$m_{0,1}$ and $m_{0,2}$, and the coupling constants $c_{N}$, $c_{M}$,
$c_{A_{N}}$, $c_{A_{M}}$, $g_{N}$, $g_{M}$, $\kappa_{1},\kappa_{2},\kappa
_{1}^{\prime}$, and $\kappa_{2}^{\prime}$. 
To determine these parameters, we use the experimental values of the
masses of the four baryonic states, the partial decay widths of the baryonic
resonances into a nucleon and a pseudoscalar meson, $\Gamma_{N(1535)\rightarrow N\pi}$, 
$\Gamma_{N(1535)\rightarrow N\eta}$, $\Gamma_{N(1650)\rightarrow N\pi}$, 
$\Gamma_{N(1650)\rightarrow N\eta}$, and
$\Gamma_{N(1440)\rightarrow N(939)\pi}$, and the axial coupling constant
$g_{A}^{N(939)}$, as well as lattice results \cite{Takahashi} for
$g_{A}^{N(1440)}$, $g_{A}^{N(1535)}$, and $g_{A}^{N(1650)}$. In total, there are thirteen
experimental values, which are fitted to twelve parameters.
The parameters $Z$, $w$,
$\varphi_{N}$, and $\varphi_{S}$ are already determined by meson physics \cite{denis}.

For the baryon masses we use the values given by the PDG \cite{PDG}. Since our
model does not contain isospin-breaking effects it is not expected to describe
the baryon masses to the (in some cases very high) experimental precision.
Therefore we assume a $5\%$ uncertainty of the masses [a strategy that was
already followed in the fit of Ref.\ \cite{denisnf3}].

The expressions for the decay widths into pseudoscalar mesons and the axial
coupling constants are given in Appendices \ref{sec:Decay Widths} and
\ref{sec:AxialCouplingConstants}. The experimental values of the decay widths
are obtained from the total width and the branching ratios given by the PDG
\cite{PDG}. The nucleon axial coupling constant is also quoted by the PDG
\cite{PDG}, while all other axial coupling constants result from lattice-QCD
calculations \cite{Takahashi}.

Using a standard $\chi^{2}$ procedure we find that three acceptable and almost
equally deep minima exist. Their corresponding parameter values are given in
Table \ref{tab:3Results}. 
\begin{table}[tbh]
\caption{The parameter values of the three $\chi^{2}$ minima and the
comparison to experimental quantities.}%
\label{tab:3Results}
\begin{center}%
\begin{tabular}
[c]{l@{\qquad}rl|rl|rl|rl}
& \multicolumn{2}{c|}{minimum 1} & \multicolumn{2}{c|}{minimum 2} &\multicolumn{2}{c|}{minimum 3} &
\multicolumn{2}{c}{experiment/lattice}\\\hline
&  &  &  &  &  & \\
$m_{0,1}~[$GeV$]$ & 	0.1393 & $\pm$ 0.0026 & 		0.14 & $\pm$ 0.11 & 		$-1.078$ &$\pm$ 0.017 & \multicolumn{2}{c}{-}\\
$m_{0,2}~[$GeV$]$ & 	$-0.2069$ & $\pm$ 0.0027 & 	$-0.18$ & $\pm$ 0.12 & 	0.894 &$\pm$ 0.019 & \multicolumn{2}{c}{-}\\
$c_{N}$ & $			-2.071$ & $\pm$ 0.023 & 		$-2.83$ & $\pm$ 0.39 & 	$-33.6$ &$\pm$ 2.2 & \multicolumn{2}{c}{-}\\
$c_{M}$ & 			12.4 & $\pm$ 1.3 & 			11.7 & $\pm$ 1.8 & 			$-19.1$ &$\pm$ 3.1 & \multicolumn{2}{c}{-}\\
$c_{A_{N}}$ & 		$-1.00$ & $\pm$ 0.23 & 		0.03 & $\pm$ 0.40 & 		$-2.68$ &$\pm$ 0.80 & \multicolumn{2}{c}{-}\\
$c_{A_{M}}$ & 		$-51.0$ & $\pm$ 2.8 & 		80 & $\pm$ 41 & 		$-71.7$ &$\pm$ 6.5 & \multicolumn{2}{c}{-}\\
$g_{N}$ & 			15.485 & $\pm$ 0.012 & 		15.24 & $\pm$ 0.36 & 		10.58 &$\pm$ 0.24 &\multicolumn{2}{c}{-}\\
$g_{M}$ & 			17.96 & $\pm$ 0.17 & 		18.26 & $\pm$ 0.52 & 		13.07 &$\pm$ 0.33 & \multicolumn{2}{c}{-}\\
$\kappa_{1}~[$GeV$^{-1}]$ & 			37.80 & $\pm$ 0.26 & 	59.9 & $\pm$ 8.5 & 	32.4 &$\pm$ 4.2 & \multicolumn{2}{c}{-}\\
$\kappa_{1}^{\prime}~[$GeV$^{-1}]$ & 57.12 & $\pm$ 0.29 & 	29.8 & $\pm$ 6.6 & 			55.2 &$\pm$ 4.0 & \multicolumn{2}{c}{-}\\
$\kappa_{2}~[$GeV$^{-1}]$ & 			$-20.7$ & $\pm$ 2.5 & 	32 & $\pm$ 13 & 	$-20$ &$\pm$  13 & \multicolumn{2}{c}{-}\\
$\kappa_{2}^{\prime}~[$GeV$^{-1}]$ & 41.5 & $\pm$ 3.2 &  		$-8$ & $\pm$ 13 & 	48.9 &$\pm$ 4.5 & \multicolumn{2}{c}{-}\\
&  &  &  &  &  & \\
$m_{N} ~[$GeV$]$    & 	0.9389 & $\pm$ 0.0010 & 		0.9389 & $\pm$ 0.0010 & 	0.9389 &$\pm$ 0.0010 &  		0.9389 & $\pm$ 0.001\\
$m_{N(1440)}~[$GeV$]$ & 	1.430  & $\pm$ 0.071 & 		1.432  & $\pm$ 0.073 & 	1.429 &$\pm$ 0.074 &  		1.43   & $\pm$ 0.07\\
$m_{N(1535)}~[$GeV$]$ & 	1.561  & $\pm$ 0.065 & 		1.585  & $\pm$ 0.069 & 	1.559 &$\pm$ 0.069 &  		1.53   & $\pm$ 0.08\\
$m_{N(1650)}~[$GeV$]$ & 	1.658  & $\pm$ 0.076 & 		1.619  & $\pm$ 0.071 & 	1.663 &$\pm$ 0.081 &  		1.65   & $\pm$ 0.08\\
$\Gamma_{N(1440)\to N\pi} ~[$GeV$]$  & 	0.195 & $\pm$ 0.087 & 	0.195 & $\pm$ 0.088 & 	0.196 &$\pm$ 0.087 & 	0.195 & $\pm$ 0.087\\
$\Gamma_{N(1535)\to N\pi} ~[$GeV$]$  & 	0.072 & $\pm$ 0.019 & 	0.073 & $\pm$ 0.019&  	0.072 &$\pm$ 0.019 & 	0.068 & $\pm$ 0.019\\
$\Gamma_{N(1535)\to N\eta} ~[$GeV$]$ &  	0.0055 & $\pm$ 0.0025 & 	0.0062 & $\pm$ 0.0024 &  	0.0055 &$\pm$ 0.0027 & 	0.063 & $\pm$ 0.018\\
$\Gamma_{N(1650)\to N\pi} ~[$GeV$]$  & 	0.112 & $\pm$ 0.033 & 	0.114 & $\pm$  0.033 & 	0.112 &$\pm$ 0.033 & 	0.105 & $\pm$ 0.037\\
$\Gamma_{N(1650)\to N\eta} ~[$GeV$]$ & 	0.0117 & $\pm$ 0.0038 & 	0.0109 & $\pm$ 0.0038 & 	0.0119 &$\pm$ 0.0038 & 	0.015 & $\pm$ 0.008\\
$g_{A}^{N}$ 			& 1.2670 & $\pm$ 0.0025 & 	1.2670  & $\pm$ 0.0025 & 		1.2670 &$\pm$ 0.0025 &  			1.267 & $\pm$ 0.003\\
$g_{A}^{N(1440)}$ 	& 1.20 	 & $\pm$ 0.20 & 		1.19  & $\pm$  0.20 & 		1.21 &$\pm$ 0.21  &  			1.2   & $\pm$ 0.2\\
$g_{A}^{N(1535)}$ 	& 0.20 & $\pm$ 0.30 & 		0.21 & $\pm$ 0.30 & 		0.20 &$\pm$ 0.31 & 				0.2   & $\pm$ 0.3\\
$g_{A}^{N(1650)}$ 	& 0.55 & $\pm$ 0.20 & 		0.55 & $\pm$ 0.20 & 		0.55 &$\pm$ 0.20 & 				0.55  & $\pm$ 0.2\\
&  &  &  &  &  & \\
$\chi^2$ &   10.3 & &  10.7 & &  10.3 & &\multicolumn{2}{c}{-}
\end{tabular}
\end{center}
\end{table}
It is interesting to note that the first two minima lead to small values of $m_{0,1}$ and $m_{0,2}$,
while the third one features values of these constants which are close to the vacuum mass of the nucleon.
Thus, for the first two minima, the main contribution to all masses arises from chiral symmetry breaking, while in
the third minimum, most of the mass is generated by another source, e.g.\ a gluon condensate.



The numerical results for the experimental quantities obtained using the above
parameters are also given in Table~\ref{tab:3Results}. Most of these
quantities are described by all solutions of the model within one standard
deviation. The most important exception is the ${N(1535)\rightarrow N\eta}$
decay width, which deviates by about an order of magnitude from the
experimental value for all scenarios explored (in fact, this deviation completely
dominates the value of $\chi^2$). Note that this
quantity was also not well described in the study of Ref.\ \cite{Gallas:2009qp}. 
Thus, including more multiplets does not solve this problem, as was erroneously speculated in that reference. 
Other ideas towards a solution are described in the next section.

It is interesting to discuss the numerical results for the mass matrix $M$ and the 
mixing matrix $U$:
\begin{itemize}
\item MINIMUM 1: Using the parameters corresponding to minimum 1 the 
mass matrix (\ref{eq:mass matrix}) reads
\[
M_{\text{min1}}=\left(
\begin{array}
[c]{rrrr}%
  0.926 & 0.071 & -0.034  & 0.173 \\
 0.071 & -1.623  & -0.173   & 0.034 \\
-0.034 & -0.173 & 1.402     & 0.228 \\
  0.173 & 0.034 & 0.228    & -1.555
\end{array}
\right)  \text{GeV}\; .
\]
Furthermore, with
the numerical value for the transformation matrix $U_{ij}$ defined in
Eq.\ (\ref{eq:U}) and composed of the eigenvectors of the mass matrix,
Eq.\ (\ref{eq:Psi phys}) can be written as
\begin{equation}
\left(
\begin{array}
[c]{c}%
N_{939}\\
\gamma^{5}N_{1535}\\
N_{1440}\\
\gamma^{5}N_{1650}%
\end{array}
\right)  =\left(
\begin{array}
[c]{rrrr}%
   \mathbf{-0.996}	& -0.025				& -0.046				& -0.074 \\
   0.075				& \mathbf{-0.492}	& 0.039				& \mathbf{-0.867} \\
   -0.050			& -0.057				& \mathbf{ 0.995}		& 0.073 \\
   0.010				& \mathbf{0.869}		& 0.086				& \mathbf{-0.488}
\end{array}
\right)  \left(
\begin{array}
[c]{c}%
\Psi_{N}\\
\gamma^{5}\Psi_{N\ast}\\
\Psi_{M}\\
\gamma^{5}\Psi_{M\ast}%
\end{array}
\right) \; . \label{mixingmatrix}%
\end{equation}
Here one can see that, to a first approximation, $N_{939}\approx\Psi_{N}$,
$N_{1440}\approx\Psi_{M}$, $N_{1535}\approx\Psi_{M\ast}$, and $N_{1650}%
\approx\Psi_{N\ast}$. Furthermore, the two negative-parity states $N_{1535}$
and $N_{1650}$ mix appreciably with each other; the mixing angle is $\sim 30^{\circ}$. 

In order to decide which states form chiral partners, we also compute the masses as a function
of $\varphi_N$, keeping $\varphi_S$ at its vacuum value. This allows us to trace the masses when
chiral symmetry is restored, $\varphi_N \rightarrow 0$. [Note that $\varphi_S$ only appears together
with a factor $\varphi_N$ in the mass matrix (\ref{eq:mass matrix})]. The result is shown in Fig.\ \ref{fig1}, from
which we unanimously conclude that $N(939)$ and $N(1535)$ are chiral partners, with a common
mass $m_{0,1} = 139$ MeV when chiral symmetry is restored. Consequently,
$N(1440)$ and $N(1650)$ are chiral partners with a mass $|m_{0,2}| = 207$ MeV as $\varphi_N \rightarrow 0$.
\begin{figure}[ht]
	\centering
		\includegraphics{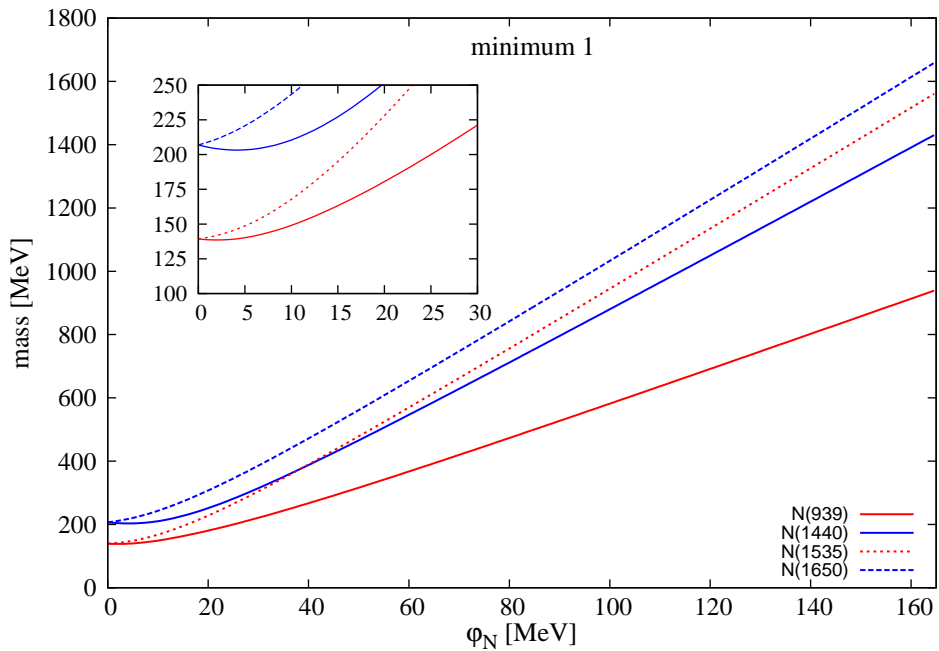}
	\caption{Masses as a function of $\varphi_N$ for minimum 1.}
	\label{fig1}
\end{figure}

\item MINIMUM 2: In this case, the mass matrix reads
\[
M_{\text{min2}}=\left(
\begin{array}
[c]{rrrr}%
	 0.925 & -0.111 & -0.017 & 0.161 \\
	-0.111 & -1.583 &-0.161  &0.017 \\
	-0.017 &-0.161 &  1.415 &-0.146 \\
	 0.161 & 0.017 &-0.146 & -1.590
\end{array}
\right)  \text{GeV}\;.
\]
Furthermore, the second minimum has the following transformation matrix:
\begin{equation}
\left(
\begin{array}
[c]{c}%
N_{939}\\
\gamma^{5}N_{1535}\\
N_{1440}\\
\gamma^{5}N_{1650}%
\end{array}
\right)  =\left(
\begin{array}
[c]{rrrr}%
	\mathbf{-0.996} 	& 0.046  			& -0.039  			&-0.061  \\
	-0.002 			& \mathbf{0.806}  	& 0.072 				& \mathbf{0.587}  \\
	-0.038 			& -0.052 			& \mathbf{0.997} 	&-0.051  \\
	0.076  			& \mathbf{0.588} 	& -0.007 			&\mathbf{-0.805} 
\end{array}
\right)  \left(
\begin{array}
[c]{c}%
\Psi_{N}\\
\gamma^{5}\Psi_{N\ast}\\
\Psi_{M}\\
\gamma^{5}\Psi_{M\ast}%
\end{array}
\right)  .
\end{equation}
As for minimum 1, the negative-parity states mix strongly but the
mixing matrix is different. Here, we may conclude that $N(1650)$ can be
predominantly assigned to $\Psi_{M\ast}$. 

In order to decide which states form chiral partners, we again compute the masses as a function
of $\varphi_N$, keeping $\varphi_S$ at its vacuum value. The result is shown in Fig.\ \ref{fig2}, from
which we again unanimously conclude that $N(939)$ and $N(1535)$ are chiral partners, with a common
mass $m_{0,1} = 144$ MeV when chiral symmetry is restored. Consequently,
$N(1440)$ and $N(1650)$ are chiral partners with a mass $|m_{0,2}| = 178$ MeV as $\varphi_N \rightarrow 0$.
\begin{figure}[ht]
	\centering
		\includegraphics{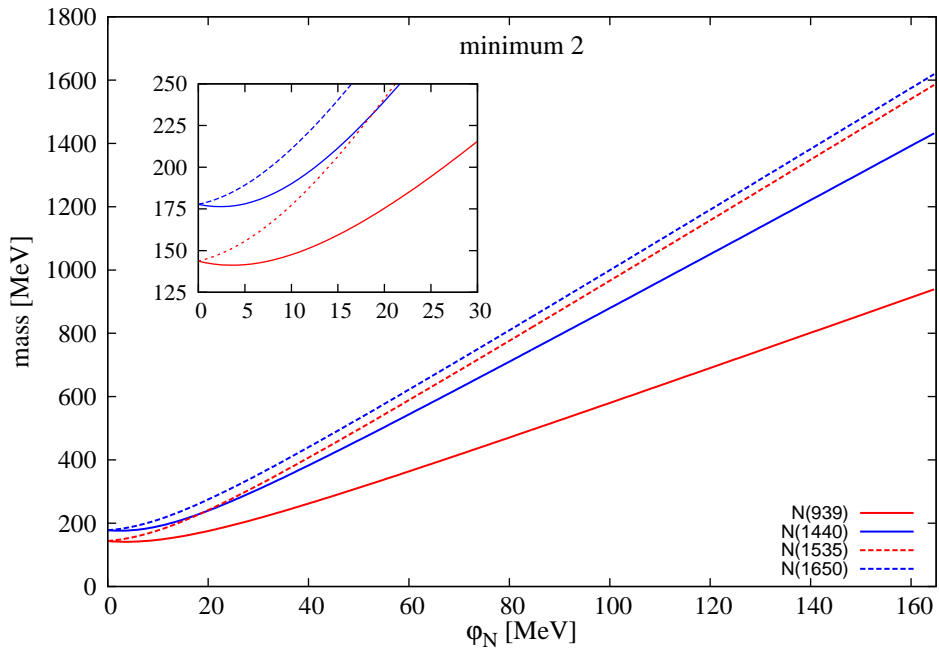}
	\caption{Masses as a function of $\varphi_N$ for minimum 2.}
	\label{fig2}
\end{figure}

\item MINIMUM 3: In this case, the mass matrix reads
\[
M_{\text{min3}}=\left(
\begin{array}
[c]{rrrr}%
  0.549 & 0.084 &-0.092 & -0.986 \\
 0.084  & -1.192  & 0.986 & 0.092 \\
-0.092  & 0.986  & 0.970 &  0.253 \\
 -0.986 & 0.092  & 0.253  & -1.181
\end{array}
\right)  \text{GeV}\;.
\]
The transformation matrix of the third minimum has a form that is completely different from those
of the other two minima:
\begin{equation}
\left(
\begin{array}[c]{c}
N_{939}\\
\gamma^{5}N_{1535}\\
N_{1440}\\
\gamma^{5}N_{1650}
\end{array}
\right)  =\left(
\begin{array}[c]{rrrr}
   \mathbf{ -0.865}	& -0.163				& \mathbf{-0.312}	& \mathbf{ 0.358} \\
   0.140				& \mathbf{ 0.830}	& \mathbf{-0.359}	& \mathbf{ 0.404}  \\
   \mathbf{-0.292}	& \mathbf{ 0.327}	& \mathbf{0.875}		& 0.207			 \\
   \mathbf{-0.384}	& \mathbf{ 0.422}	& -0.093				& \mathbf{-0.816}
\end{array} 
\right)  \left(
\begin{array}[c]{c}
\Psi_{N}\\
\gamma^{5}\Psi_{N\ast}\\
\Psi_{M}\\
\gamma^{5}\Psi_{M\ast}
\end{array}
\right) \;.
\end{equation}
In this case, all states mix strongly with each other.

In order to decide which states form chiral partners, we again compute the masses as a function
of $\varphi_N$, keeping $\varphi_S$ at its vacuum value. The result is shown in Fig.\ \ref{fig3}, from
which we again unanimously conclude that $N(939)$ and $N(1535)$ are chiral partners, with a common
mass $m_{0,2} = 894$ MeV when chiral symmetry is restored. Consequently,
$N(1440)$ and $N(1650)$ are chiral partners with a mass $|m_{0,1}| = 1078$ MeV as $\varphi_N \rightarrow 0$.
\begin{figure}[ht]
	\centering
		\includegraphics{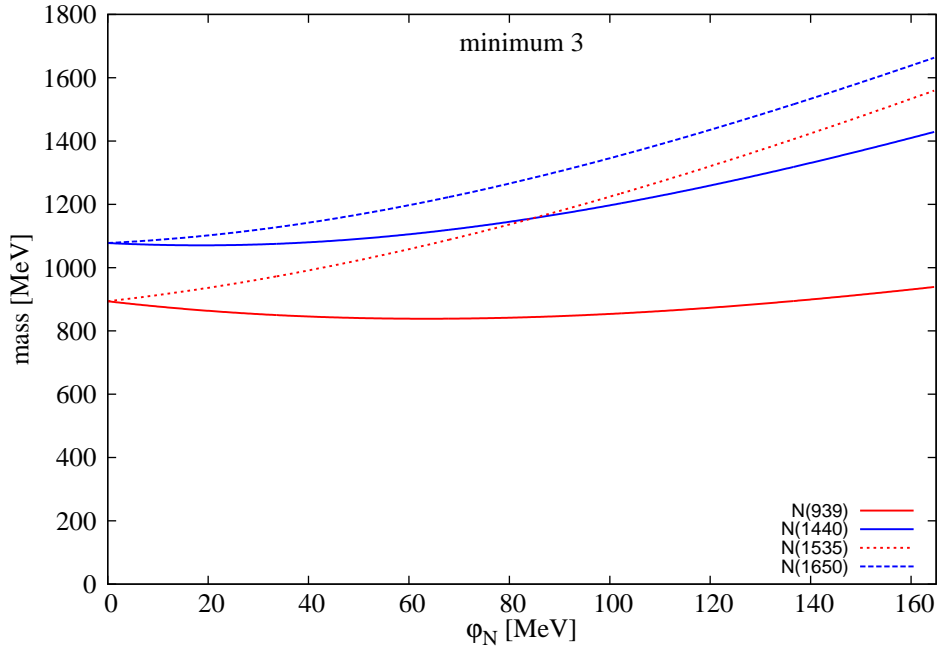}
	\caption{Masses as a function of $\varphi_N$ for minimum 3.}
	\label{fig3}
\end{figure}

\end{itemize}

\section{Conclusions and Outlook}

In this work we have studied the generalization of the eLSM to the three-flavor
case, thus including baryons with strangeness ($N_{f}=3$).
We have found that, in a chiral quark-diquark model for the baryons, we
naturally need to consider four baryonic multiplets, 
if we require the presence of chirally invariant mass terms like in the mirror assignment. 
Subsequently, we have reduced the model to the case $N_{f}=2$ and
performed a fit of the parameters of the model to the masses
and decay widths as well as the axial coupling constants of the nucleonic resonances $N(939),$
$N(1440),$ $N(1535),$ and $N(1650)$. Masses and decay widths as well as the
axial coupling constant of the nucleon are experimentally known  \cite{PDG}, for the axial coupling constants
of the other resonances we used lattice-QCD data \cite{Takahashi}.

From this fit, we found three minima which,
with the exception of the decay $N(1535) \rightarrow N \eta$, 
yield results for the masses, for the decay widths, and for the axial coupling
constants that are in very good agreement with data, see Table 1. 
Studying the approach to chiral-symmetry restoration $\varphi_N \rightarrow 0$, we were able to unanimously identify
which of the four nucleonic resonances form chiral partners. For all three minima, these
are the pairs $N(939),\, N(1535)$, as well as $N(1440),\, N(1650)$.

Finally, let us discuss the issue with the decay width 
$N(1535)\rightarrow N\eta$. Our result that the theoretical value turns 
out to be too small when compared to the
experimental value is stable under parameter variations. This
implies that further studies are needed to understand the resonance $N(1535).$
Some authors have argued that $N(1535)$ may contain a sizable amount of
$s\bar{s}$ \cite{sheng,liu,chao}. Another interesting possibility is
to investigate the role of the chiral anomaly in the baryonic sector
\cite{psgproc}, which can lead to an enhanced coupling to the resonances
$\eta$ and $\eta^{\prime}$.

In the very recent study of Ref.\ \cite{harada} a chiral baryonic model with
three flavors was constructed by making use of parity doublets. There, a large
variety of baryonic fields was included (also the decuplet is present), but no
(axial-)vector degrees of freedom were considered in the mesonic sector. The chirally
invariant contribution to the nucleon mass is in the range $500 - 800$ MeV, in agreement with
our result for minimum 3. Interestingly, in Ref.\ \cite{harada} upper bounds for
the axial coupling constants were derived which fit well to our results.

Finally, in order to decide
which of the three minima that resulted from our fit is preferable, we plan to investigate the
complete three-flavor case. Note that most of 
the parameters of the Lagrangian (\ref{eq:LagNf=3}) are already determined from
the $N_{f}=2$ fit,
but many more experimental data, such as the masses of hyperons and their decay widths, 
are available to discriminate between the three minima. 
The obtained values for the coupling constants of hyperons to (pseudo)scalar
and (axial-)vector mesons will be relevant for studies of
scattering processes in the vacuum
\cite{pptopkaon,pptoppphi,pptoppetaprime,kaonp} as well as for neutron stars
\cite{hyperonstar,giuseppe}. In connection to the latter topic, one can study
nuclear matter at nonzero density and inhomogeneous chiral condensation, thus
extending previous investigations on the subjects \cite{achim,pagliara} in a
more complete framework.\\[0.2cm]

\noindent \textbf{Acknowledgements:\newline} We thank S.\ Gallas, G.\ Pagliara, P.\
Kov\'{a}cs, and Gy.\ Wolf for useful discussions. L.O.\ acknowledges support by HGS-HIRe/HQM. M.Z. was supported by the Hungarian OTKA Fund No. K109462 and HIC for FAIR.

\appendix

\section{Mass Degeneracy in the Case of an $N_{f}=2$ Lagrangian without
$\kappa$ terms}

\label{sec:masses without kappas} 

In this Appendix we want to clarify why it
is mandatory to include the $\kappa$ and $\epsilon$ terms in the
Lagrangian (\ref{eq:LagNf=3}), although they are not dilatation-invariant.
Therefore, we consider the two-flavor case, given in Eq.\ (\ref{eq:LagNf=2})
and in Appendix \ref{sec:Full Nf=2 Lag}. Setting the constants $\kappa
_{1},\kappa_{2},\kappa_{1}^{\prime},\kappa_{2}^{\prime}$ (and $\epsilon_{i}$
with $i=1,2,3,4$) to zero, the part of the Lagrangian which contains the terms contributing to the mass 
matrix of the four fields $\Psi_{N}$, $\Psi_{N\ast}$, $\Psi_{M}$, and $\Psi_{M\ast}$ reads
$\mathcal{L}_{\text{mass}}=-\bar{\Psi}M^{\prime}\Psi$. 
The definition of the vector $\Psi$ is given in Eq.\ (\ref{eq:Psi}) and the mass
matrix is given by
\[
M^{\prime}=\frac{1}{2}\left(
\begin{array}
[c]{cccc}%
g_{N}\varphi_{N} & 0 & m_{0,1}+m_{0,2} & m_{0,1}-m_{0,2}\\
0 & -g_{N}\varphi_{N} & m_{0,2}-m_{0,1} & -m_{0,1}-m_{0,2}\\
m_{0,1}+m_{0,2} & m_{0,2}-m_{0,1} & g_{M}\varphi_{N} & 0\\
m_{0,1}-m_{0,2} & -m_{0,1}-m_{0,2} & 0 & -g_{M}\varphi_{N}%
\end{array}
\right) \;.
\]
Evaluating $\det\bigl(M^{\prime}-m_{i}\mathbbm{1}_{4\times4}\bigr)=0$ and denoting 
$\Omega_{1/2}    = \sqrt{\frac{1}{4}(m_{0,1}\pm m_{0,2})^2 + \frac{1}{16}(g_N \mp g_M)^2 \varphi_N^2}$, 
we find the eigenvalues
\begin{align}
m_1 &= \Omega_1 + \Omega_2 = - m_2\;,\n
m_3 &= \Omega_1 - \Omega_2 = - m_4\;.
\label{eq:appendix eigenvalues}%
\end{align}
Due to the definition of the vectors $\Psi$ given in Eq.\ (\ref{eq:Psi}) and
the definition of physical states in Eq.\ (\ref{eq:Psi phys}), the physical
masses correspond to the eigenvalues as follows: $m_{939}=m_{1}$ and
$m_{1535}=-m_{2}$ as well as $m_{1440}=m_{3}$ and $m_{1650}=-m_{4}$. Having
this in mind and considering the results for the eigenvalues given in Eq.\
(\ref{eq:appendix eigenvalues}) it is clear that the masses of $N(939)$ and
$N(1535)$ as well as the masses of $N(1440)$ and $N(1650)$ would be
degenerate. The only possibility to avoid a mass degeneracy but still
keep chiral symmetry is to introduce the $\kappa$ (and $\epsilon$) terms as
shown in Eq.\ (\ref{eq:LagNf=3}).

\newpage 

\section{Explicit Lagrangian for $N_{f}=3$ in Terms of Parity Eigenstates}

\label{sec:Full Nf=3 Lag}

From Eq.\ (\ref{eq:def of B fields}) and the Lagrangian (\ref{eq:LagNf=3}) one
obtains the following baryonic Lagrangian for $N_{f}=3$ flavors as a function
of parity eigenstates:%
\begin{align*}
\mathcal{L}_{N_f=3}=  &  \Tr\bigl\{
\bar{B}_{NR}i\gamma_{\mu} D_{NR}^{\mu} B_{NR} +
\bar{B}_{NL}i\gamma_{\mu} D_{NL}^{\mu} B_{NL} +
\bar{B}_{N\ast R}i\gamma_{\mu} D_{NR}^{\mu} B_{N\ast R} + 
\bar{B}_{N\ast L}i\gamma_{\mu} D_{NL}^{\mu} B_{N\ast L}  \\
& \quad +
\bar{B}_{MR}i\gamma_{\mu} D_{ML}^{\mu} B_{MR} +
\bar{B}_{ML}i\gamma_{\mu} D_{MR}^{\mu} B_{ML} +
\bar{B}_{M\ast R}i\gamma_{\mu} D_{ML}^{\mu} B_{M\ast R} +
\bar{B}_{M\ast L}i\gamma_{\mu} D_{MR}^{\mu} B_{M\ast L}\bigr\} \\
&  \quad+c_{A_{N}}\Tr\bigl\{
\bar{B}_{NR}i\gamma_{\mu} R^{\mu} B_{N\ast R} +
\bar{B}_{N\ast R}i\gamma_{\mu} R^{\mu} B_{NR} -
\bar{B}_{NL}i\gamma_{\mu} L^{\mu} B_{N\ast L} -
\bar{B}_{N\ast L}i\gamma_{\mu} L^{\mu} B_{NL} \bigr\} \\
&  \quad+c_{A_{M}}\Tr\bigl\{
\bar{B}_{MR}i\gamma_{\mu} L^{\mu} B_{M\ast R} +
\bar{B}_{M\ast R}i\gamma_{\mu} L^{\mu} B_{MR} -
\bar{B}_{ML}i\gamma_{\mu} R^{\mu} B_{M\ast L} -
\bar{B}_{M\ast L}i\gamma_{\mu} R^{\mu} B_{ML} \bigr\} \\
&  \quad- g_{N}\Tr\bigl\{
\bar{B}_{NL} \Phi B_{NR}+
\bar{B}_{NR} \Phi^{\dagger} B_{NL}+
\bar{B}_{N\ast L} \Phi B_{N\ast R}+
\bar{B}_{N\ast L} \Phi^{\dagger} B_{N\ast R} \bigr\}\\
&  \quad- g_{M}\Tr\bigl\{
\bar{B}_{ML} \Phi^{\dagger} B_{MR}+
\bar{B}_{MR} \Phi B_{ML}+
\bar{B}_{M\ast L} \Phi^{\dagger} B_{M\ast R}+
\bar{B}_{M\ast L} \Phi B_{M\ast R} \bigr\}\\
&  \quad-\frac{\kappa_{1}}{2}\Tr\bigl\{-
\bar{B}_{NL} \Phi B_{NR} \Phi^{\dagger}-
\bar{B}_{NR} \Phi^{\dagger} B_{NL} \Phi +
\bar{B}_{N\ast L} \Phi B_{N\ast R} \Phi^{\dagger}+
\bar{B}_{N\ast R} \Phi^{\dagger} B_{N\ast L} \Phi\\
& \qquad \qquad \quad -
\bar{B}_{NL} \Phi B_{N\ast R} \Phi^{\dagger}+
\bar{B}_{NR} \Phi^{\dagger} B_{N\ast L} \Phi +
\bar{B}_{N\ast L} \Phi B_{NR} \Phi^{\dagger}-
\bar{B}_{N\ast R} \Phi^{\dagger} B_{NL} \Phi  \bigr\}\\
&  \quad-\frac{\kappa_{1}^{\prime}}{2}\Tr\bigl\{-
\bar{B}_{NL} \Phi B_{NR} \Phi-
\bar{B}_{NR} \Phi^{\dagger} B_{NL} \Phi^{\dagger} +
\bar{B}_{N\ast L} \Phi B_{N\ast R} \Phi+
\bar{B}_{N\ast R} \Phi^{\dagger} B_{N\ast L}\Phi^{\dagger} \\
& \qquad \qquad \quad +
\bar{B}_{NL} \Phi B_{N\ast R} \Phi-
\bar{B}_{NR} \Phi^{\dagger} B_{N\ast L} \Phi^{\dagger}-
\bar{B}_{N\ast L} \Phi B_{NR} \Phi+
\bar{B}_{N\ast R} \Phi^{\dagger} B_{NL} \Phi^{\dagger}  \bigr\}\\
&  \quad-\frac{\kappa_{2}}{2}\Tr\bigl\{-
\bar{B}_{ML} \Phi^{\dagger} B_{MR} \Phi^{\dagger}-
\bar{B}_{MR} \Phi B_{ML} \Phi +
\bar{B}_{M\ast L} \Phi^{\dagger} B_{M\ast R} \Phi^{\dagger}+
\bar{B}_{M\ast R} \Phi B_{M\ast L} \Phi\\
& \qquad \qquad \quad -
\bar{B}_{ML} \Phi^{\dagger} B_{M\ast R} \Phi^{\dagger}+
\bar{B}_{MR} \Phi B_{M\ast L} \Phi +
\bar{B}_{M\ast L} \Phi^{\dagger} B_{MR} \Phi^{\dagger} -
\bar{B}_{M\ast R} \Phi B_{ML} \Phi \bigr\}\\
&  \quad-\frac{\kappa_{2}^{\prime}}{2}\Tr\bigl\{-
\bar{B}_{ML} \Phi^{\dagger} B_{MR} \Phi-
\bar{B}_{MR} \Phi B_{ML}  \Phi^{\dagger} +
\bar{B}_{M\ast L} \Phi^{\dagger} B_{M\ast R} \Phi+
\bar{B}_{M\ast R} \Phi B_{M\ast L} \Phi^{\dagger}\\
& \qquad \qquad \quad +
\bar{B}_{ML} \Phi^{\dagger} B_{M\ast R} \Phi -
\bar{B}_{MR} \Phi B_{M\ast L} \Phi^{\dagger}-
\bar{B}_{M\ast L} \Phi^{\dagger} B_{MR} \Phi +
\bar{B}_{M\ast R} \Phi B_{ML} \Phi^{\dagger} \bigr\}\\
&  \quad-\frac{m_{0,1} + m_{0,2}}{2}\Tr\Bigl\{
\bar{B}_{NL}B_{MR}+
\bar{B}_{NR}B_{ML}+
\bar{B}_{N\ast L}B_{M\ast R}+
\bar{B}_{N\ast R}B_{M\ast L} \n
& \qquad \qquad \qquad \qquad ~ +  
\bar{B}_{ML}B_{NR}+
\bar{B}_{MR}B_{NL}+
\bar{B}_{M\ast L}B_{N\ast R}+
\bar{B}_{M\ast R}B_{N\ast L}
\Bigr\}\\
&  \quad-\frac{m_{0,1}-m_{0,2}}{2}\Tr\Bigl\{
\bar{B}_{NL}B_{M\ast R}-
\bar{B}_{NR}B_{M\ast L}-
\bar{B}_{ML}B_{N\ast R}+
\bar{B}_{MR}B_{N\ast L}\n
& \qquad \qquad \qquad \qquad ~ -
\bar{B}_{N\ast L}B_{MR}+
\bar{B}_{N\ast R}B_{ML}+
\bar{B}_{M\ast L}B_{NR}-
\bar{B}_{M\ast R}B_{NL}
\Bigr\} \; ,
\end{align*}
where the covariant derivatives
\[
D_{NR}^{\mu} =\partial^{\mu}-ic_{N}R^{\mu}~\text{,}\qquad D_{NL}^{\mu}
=\partial^{\mu}-ic_{N}L^{\mu}\;,%
\]
\[
D_{MR}^{\mu} =\partial^{\mu}-ic_{M}R^{\mu}~\text{,}\qquad D_{ML}^{\mu}
=\partial^{\mu}-ic_{M}L^{\mu}\;,%
\]
with
\[
c_{N}= \frac{c_{1}+c_{2}}{2} \quad\text{ and } \quad c_{M}=\frac{c_{3}+c_{4}%
}{2}.
\]
These two constants parametrize the coupling between baryons of equal parity.
The constants
\[
c_{A_{N}}=\frac{c_{1}-c_{2}}{2} \quad\text{ and } \quad c_{A_{M}}=\frac
{c_{3}-c_{4}}{2}%
\]
describe the coupling of two baryons with different parity to (axial-)vector
mesons. The interaction of the baryonic fields with the scalar and
pseudoscalar mesonic fields are parametrized by $g_{N}$ and $g_{M}$. The
chirally invariant mass terms are characterized by $m_{0,1}$ and $m_{0,2}$.
The terms proportional to $\kappa_{1(2)}^{(\prime)}$ are introduced to avoid
mass degeneracy (see Appendix A). In total the Lagrangian has twelve free parameters.
%
\newpage

\section{Explicit Lagrangian for $N_{f}=2$ after SSB}

\label{sec:Full Nf=2 Lag} After SSB in the meson sector ($\sigma
_{N}\rightarrow\sigma_{N}+\varphi_{N}$ and $\sigma_{S}\rightarrow\sigma
_{S}+\varphi_{S}$), the full Lagrangian with two flavors describing the
nucleon, $N(1440)$, and their chiral partners, as well as their interaction
with scalar, pseudoscalar, vector, and axial-vector mesons reads
\begin{align}
\mathcal{L}=  &  \bar{\Psi}_{N}i\gamma^{\mu}\partial_{\mu}\Psi_{N}+\bar{\Psi
}_{N\ast}i\gamma^{\mu}\partial_{\mu}\Psi_{N\ast}+\bar{\Psi}_{M}i\gamma^{\mu
}\partial_{\mu}\Psi_{M}+\bar{\Psi}_{M\ast}i\gamma^{\mu}\partial_{\mu}%
\Psi_{M\ast}\nonumber\\
&  +c_{N}\Bigl(\bar{\Psi}_{N~}\gamma_{\mu}\left\{  \left[  \omega^{\mu}%
-\gamma^{5}\left(  f_{1}^{\mu}+Zw\partial^{\mu}\eta_{N}\right)  \right]
T^{0}+\left[  \boldsymbol{\rho}^{\mu}-\gamma^{5}\left(  \boldsymbol{a}%
_{1}^{\mu}+Zw\partial^{\mu}\boldsymbol{\pi}\right)  \right]  \cdot
\boldsymbol{T}\right\}  \Psi_{N}\nonumber\\
&  \qquad+\bar{\Psi}_{N\ast}\gamma_{\mu}\left\{  \left[  \omega^{\mu}%
-\gamma^{5}\left(  f_{1}^{\mu}+Zw\partial^{\mu}\eta_{N}\right)  \right]
T^{0}+\left[  \boldsymbol{\rho}^{\mu}-\gamma^{5}\left(  \boldsymbol{a}%
_{1}^{\mu}+Zw\partial^{\mu}\boldsymbol{\pi}\right)  \right]  \cdot
\boldsymbol{T}\right\}  \Psi_{N\ast}\Bigr)\nonumber\\
&  +c_{M}\Bigl(\bar{\Psi}_{M~}\gamma_{\mu}\left\{  \left[  \omega^{\mu}%
+\gamma^{5}\left(  f_{1}^{\mu}+Zw\partial^{\mu}\eta_{N}\right)  \right]
T^{0}+\left[  \boldsymbol{\rho}^{\mu}+\gamma^{5}\left(  \boldsymbol{a}%
_{1}^{\mu}+Zw\partial^{\mu}\boldsymbol{\pi}\right)  \right]  \cdot
\boldsymbol{T}\right\}  \Psi_{M}\nonumber\\
&  \qquad+\bar{\Psi}_{M\ast}\gamma_{\mu}\left\{  \left[  \omega^{\mu}%
+\gamma^{5}\left(  f_{1}^{\mu}+Zw\partial^{\mu}\eta_{N}\right)  \right]
T^{0}+\left[  \boldsymbol{\rho}^{\mu}+\gamma^{5}\left(  \boldsymbol{a}%
_{1}^{\mu}+Zw\partial^{\mu}\boldsymbol{\pi}\right)  \right]  \cdot
\boldsymbol{T}\right\}  \Psi_{M\ast}\Bigr)\nonumber\\
&  +c_{A_{N}}\Bigl\{\bar{\Psi}_{N~}\gamma_{\mu}\left[  \left(  -f_{1}^{\mu
}-Zw\partial^{\mu}\eta_{N}+\gamma^{5}\omega^{\mu}\right)  T^{0}+\left(
-\boldsymbol{a}_{1}^{\mu}-Zw\partial^{\mu}\boldsymbol{\pi}+\gamma
^{5}\boldsymbol{\rho}^{\mu}\right)  \cdot\boldsymbol{T}\right]  \Psi_{N\ast
}\nonumber\\
&  \qquad+\bar{\Psi}_{N\ast}\gamma_{\mu}\left[  \left(  -f_{1}^{\mu
}-Zw\partial^{\mu}\eta_{N}+\gamma^{5}\omega^{\mu}\right)  T^{0}+\left(
-\boldsymbol{a}_{1}^{\mu}-Zw\partial^{\mu}\boldsymbol{\pi}+\gamma
^{5}\boldsymbol{\rho}^{\mu}\right)  \cdot\boldsymbol{T}\right]  \Psi
_{N}\Bigr\}\nonumber\\
&  +c_{A_{M}}\Bigl\{\bar{\Psi}_{M}\gamma_{\mu}\left[  \left(  f_{1}^{\mu
}+Zw\partial^{\mu}\eta_{N}+\gamma^{5}\omega^{\mu}\right)  T^{0}+\left(
\boldsymbol{a}_{1}^{\mu}+Zw\partial^{\mu}\boldsymbol{\pi}+\gamma
^{5}\boldsymbol{\rho}^{\mu}\right)  \cdot\boldsymbol{T}\right]  \Psi_{M\ast
}\nonumber\\
&  \qquad+\bar{\Psi}_{M\ast}\gamma_{\mu}\left[  \left(  f_{1}^{\mu}%
+Zw\partial^{\mu}\eta_{N}+\gamma^{5}\omega^{\mu}\right)  T^{0}+\left(
\boldsymbol{a}_{1}^{\mu}+Zw\partial^{\mu}\boldsymbol{\pi}+\gamma
^{5}\boldsymbol{\rho}^{\mu}\right)  \cdot\boldsymbol{T}\right]  \Psi
_{M}\Bigr\}\nonumber\\
&  -g_{N}\Bigl\{\bar{\Psi}_{N}\left[  \left(  \sigma+\varphi_{N}+i\gamma
^{5}Z\eta_{N}\right)  T^{0}+\left(  \boldsymbol{a}_{0}+i\gamma^{5}%
Z\boldsymbol{\pi}\right)  \cdot\boldsymbol{T}\right]  \Psi_{N}\nonumber\\
&  \qquad+\bar{\Psi}_{N\ast}\left[  \left(  \sigma+\varphi_{N}+i\gamma
^{5}Z\eta_{N}\right)  T^{0}+\left(  \boldsymbol{a}_{0}+i\gamma^{5}%
Z\boldsymbol{\pi}\right)  \cdot\boldsymbol{T}\right]  \Psi_{N\ast
}\Bigr\}\nonumber\\
&  -g_{M}\Bigl\{\bar{\Psi}_{M}\left[  \left(  \sigma+\varphi_{N}-i\gamma
^{5}Z\eta_{N}\right)  T^{0}+\left(  \boldsymbol{a}_{0}-i\gamma^{5}%
Z\boldsymbol{\pi}\right)  \cdot\boldsymbol{T}\right]  \Psi_{M}\nonumber\\
&  \qquad+\bar{\Psi}_{M\ast}\left[  \left(  \sigma+\varphi_{N}-i\gamma
^{5}Z\eta_{N}\right)  T^{0}+\left(  \boldsymbol{a}_{0}-i\gamma^{5}%
Z\boldsymbol{\pi}\right)  \cdot\boldsymbol{T}\right]  \Psi_{M\ast
}\Bigr\}\nonumber\\
&  -\frac{\kappa_{1}^{\prime}+\kappa_{1}}{2\sqrt{2}}\varphi_{S}\Bigl\{-\bar
{\Psi}_{N}\left[  \left(  \sigma+\varphi_{N}+i\gamma^{5}Z\eta_{N}\right)
T^{0}+\left(  \boldsymbol{a}_{0}+i\gamma^{5}Z\boldsymbol{\pi}\right)
\cdot\boldsymbol{T}\right]  \Psi_{N}\nonumber\\
&  ~\qquad\qquad\qquad+\bar{\Psi}_{N\ast}\left[  \left(  \sigma+\varphi
_{N}+i\gamma^{5}Z\eta_{N}\right)  T^{0}+\left(  \boldsymbol{a}_{0}+i\gamma
^{5}Z\boldsymbol{\pi}\right)  \cdot\boldsymbol{T}\right]  \Psi_{N\ast
}\Bigr\}\nonumber\\
&  -\frac{\kappa_{1}^{\prime}-\kappa_{1}}{2\sqrt{2}}\varphi_{S}\Bigl\{~~\bar
{\Psi}_{N}\left[  \left(  iZ\eta_{N}+\gamma^{5}(\sigma+\varphi_{N})\right)
T^{0}+\left(  iZ\boldsymbol{\pi}+\gamma^{5}\boldsymbol{a}_{0}\right)
\cdot\boldsymbol{T}\right]  \Psi_{N\ast}\nonumber\\
&  ~\qquad\qquad\qquad-\bar{\Psi}_{N\ast}\left[  \left(  iZ\eta_{N}+\gamma
^{5}(\sigma+\varphi_{N})\right)  T^{0}+\left(  iZ\boldsymbol{\pi}+\gamma
^{5}\boldsymbol{a}_{0}\right)  \cdot\boldsymbol{T}\right]  \Psi_{N}%
\Bigr\}\nonumber\\
&  -\frac{\kappa_{2}^{\prime}+\kappa_{2}}{2\sqrt{2}}\varphi_{S}\Bigl\{-\bar
{\Psi}_{M}\left[  \left(  \sigma+\varphi_{N}-i\gamma^{5}Z\eta_{N}\right)
T^{0}+\left(  \boldsymbol{a}_{0}-i\gamma^{5}Z\boldsymbol{\pi}\right)
\cdot\boldsymbol{T}\right]  \Psi_{M}\nonumber\\
&  ~\qquad\qquad\qquad+\bar{\Psi}_{M\ast}\left[  \left(  \sigma+\varphi
_{N}-i\gamma^{5}Z\eta_{N}\right)  T^{0}+\left(  \boldsymbol{a}_{0}-i\gamma
^{5}Z\boldsymbol{\pi}\right)  \cdot\boldsymbol{T}\right]  \Psi_{M\ast
}\Bigr\}\nonumber\\
&  -\frac{\kappa_{2}^{\prime}-\kappa_{2}}{2\sqrt{2}}\varphi_{S}\Bigl\{-\bar
{\Psi}_{M}\left[  \left(  iZ\eta_{N}-\gamma^{5}(\sigma+\varphi_{N})\right)
T^{0}+\left(  iZ\boldsymbol{\pi}-\gamma^{5}\boldsymbol{a}_{0}\right)
\cdot\boldsymbol{T}\right]  \Psi_{M\ast}\nonumber\\
&  ~\qquad\qquad\qquad+\bar{\Psi}_{M\ast}\left[  \left(  iZ\eta_{N}-\gamma
^{5}(\sigma+\varphi_{N})\right)  T^{0}+\left(  iZ\boldsymbol{\pi}-\gamma
^{5}\boldsymbol{a}_{0}\right)  \cdot\boldsymbol{T}\right]  \Psi_{M}%
\Bigr\}\nonumber\\
&  -\frac{m_{0,1}+m_{0,2}}{2}\Bigl(\bar{\Psi}_{N}\Psi_{M}+\bar{\Psi}_{N\ast
}\Psi_{M\ast}+\bar{\Psi}_{M}\Psi_{N}+\bar{\Psi}_{M\ast}\Psi_{N\ast
}\Bigr)\nonumber\\
&  -\frac{m_{0,2}-m_{0,1}}{2}\Bigl(\bar{\Psi}_{N}\gamma^{5}\Psi_{M\ast}%
+\bar{\Psi}_{N\ast}\gamma^{5}\Psi_{M}-\bar{\Psi}_{M}\gamma^{5}\Psi_{N\ast
}-\bar{\Psi}_{M\ast}\gamma^{5}\Psi_{N}\Bigr)\text{.} \label{eq:full Lag Nf=2}%
\end{align}
where the coupling to (axial-)vector mesons of two baryons with equal parity
and a vector meson is parametrized by $c_{N}=(c_{1}+c_{2})/2$ and
$c_{M}=(c_{3}+c_{4})/2$ and of two baryons with opposite parity by $c_{A_{N}%
}=(c_{1}-c_{2})/2$ and $c_{A_{M}}=(c_{3}-c_{4})/2$. All other constants
are the same as in the Lagrangian (\ref{eq:LagNf=2}). The factor $w$ is introduced
due to the shift of the axial-vector fields in order to eliminate the mixing with the pseudoscalar fields,
which occurs after
SSB, and $Z$ is the so-called wave-function renormalization factor that takes care of
the normalization of the kinetic terms of the pseudoscalar mesonic fields
after the shift, see Ref.\ \cite{denis} for more details.


\newpage

\section{Decay Widths}

\label{sec:Decay Widths} 

Because of the existing experimental data \cite{PDG}, we are especially interested in the decays of nucleon
resonances into the pseudoscalar mesons $\pi$ and $\eta$. The Lagrangian
describing the decay of a resonance $N^{\ast}$ into a nucleon $N$ and a
pseudoscalar meson $P=\pi,\eta$ has the general structure
\begin{equation}
\mathcal{L}=g^{N_{\ast}\rightarrow N\partial P}\bar{N}\Gamma\gamma_{\mu
}N_{\ast}\partial^{\mu}P-ig^{N_{\ast}\rightarrow NP}\bar{N}\Gamma\gamma_{\mu
}N_{\ast}P,
\end{equation}
where $\Gamma=\gamma_{5}$ $(1)$ for a positive-(negative-)parity $N_{\ast}$.
The explicit expressions for the coupling constants $g^{N_{\ast}\rightarrow
N\partial P}$ and $g^{N_{\ast}\rightarrow NP}$ can be obtained from the
relevant terms of the Lagrangian (\ref{eq:full Lag Nf=2}), carrying out the
transformation (\ref{eq:Psi phys}). Using this the tree-level decay width can
be calculated to be
\begin{equation}
\Gamma_{N_{\ast}\rightarrow NP}=\lambda_{P}\frac{p_{f}}{8\pi m_{N_{\ast}}^{2}%
}\overline{\left\vert i\mathcal{M}\right\vert }^{2}=\kappa_{P}\frac{p_{f}%
}{4\pi m_{N_{\ast}}}\left[  g^{N_{\ast}\rightarrow NP}-(m_{N_{\ast}}\pm
m_{N})g^{N_{\ast}\rightarrow N\partial P}\right]  ^{2}(E_{N}\mp m_{N})\; ,
\end{equation}
where the upper (lower) sign is valid for a positive-(negative-)parity
$N_{\ast}$, $E_{N}$ is the nucleon energy in the rest frame of the decaying
$N_{\ast}$, while the magnitude of the three-momenta of the decay products is
\begin{equation}
p_{f}=\frac{1}{2m_{N_{\ast}}}\sqrt{(m_{N_{\ast}}^{2}-m_{N}^{2}-m_{P}^{2}%
)^{2}-4m_{N}^{2}m_{P}^{2}}.
\end{equation}
Furthermore the factor $\lambda_{P}$ is added by hand and should
\begin{itemize}
\item for $P = \pi$ pay attention to the three possible isospin states of the
pion, i.e.,
\begin{align*}
\lambda_{\pi} = 3.
\end{align*}
\item and for $P=\eta$ take into account that
\[
\eta=\eta_{N}\cos\phi_{P}+\eta_{S}\sin\phi_{P},
\]
where $\eta_{N}\equiv(\bar{u}u+\bar{d}d)/\sqrt{2}$ and $\eta_{S}\equiv\bar
{s}s$ and $\phi_{P}$ is the mixing angle. Its value lies between $-32^{\circ}$
and $-45^{\circ}$ \cite{phiP}. In this paper we have chosen $\phi
_{P}=-44.6^{\circ}$ obtained from Ref.\ \cite{denisnf3}.  It is
assumed that the amplitude of the decay $N_{\ast}\rightarrow N\eta_{S}$ is
massively suppressed. This means that to good approximation
\[
\Gamma_{N_{\ast}\rightarrow N\eta}\simeq\cos^{2}\phi_{P}\Gamma_{N_{\ast
}\rightarrow N\eta_{N}}.
\]
Thus:
\[
\lambda_{\eta}=\cos^{2}\phi_{P}\text{ }.
\]

\end{itemize}


\section{Axial Coupling Constants}

\label{sec:AxialCouplingConstants} 

The Lagrangians in the Appendix
\ref{sec:Full Nf=3 Lag} and Appendix \ref{sec:Full Nf=2 Lag} are invariant
under $U_{A}=\text{exp}(-i\theta^{a}\gamma^{5}\tau^{a}/2)\in U(N_{f})_{A}$
axial transformations ($\theta^{a}$ are the parameters and $\tau^{a}/2$ the
generators). Due to Noether's theorem \cite{noether} one gets the following
axial current
\begin{align}
A^{a,\mu}=  &  g_{A}^{(1)}\bar{\Psi}_{N}\gamma^{\mu}\gamma^{5}\frac{\tau^{a}%
}{2}\Psi_{N}+g_{A}^{(1)}\bar{\Psi}_{N\ast}\gamma^{\mu}\gamma^{5}\frac{\tau
^{a}}{2}\Psi_{N\ast}+g_{A}^{(2)}\bar{\Psi}_{M}\gamma^{\mu}\gamma^{5}\frac
{\tau^{a}}{2}\Psi_{M}+g_{A}^{(2)}\bar{\Psi}_{M\ast}\gamma^{\mu}\gamma^{5}%
\frac{\tau^{a}}{2}\Psi_{M\ast}\nonumber\\
&  +g_{A}^{(12)}\bar{\Psi}_{N}\gamma^{\mu}\frac{\tau^{a}}{2}\Psi_{N\ast}%
+g_{A}^{(12)}\bar{\Psi}_{N\ast}\gamma^{\mu}\frac{\tau^{a}}{2}\Psi_{N}%
+g_{A}^{(34)}\bar{\Psi}_{M}\gamma^{\mu}\frac{\tau^{a}}{2}\Psi_{M\ast}%
+g_{A}^{(34)}\bar{\Psi}_{M\ast}\gamma^{\mu}\frac{\tau^{a}}{2}\Psi_{M}\;,
\label{eq:axial current}%
\end{align}
where
\[
g_{A}^{(1)}=1-\frac{c_{N}}{g_{1}}\left(  1-\frac{1}{Z^{2}}\right)  \;, \quad
g_{A}^{(2)}=-1+\frac{c_{M}}{g_{1}}\left(  1-\frac{1}{Z^{2}}\right)  \;,
\]
are the axial coupling constants of the bare fields $\Psi_{N}$, $\Psi_{N\ast}%
$, $\Psi_{M}$, and $\Psi_{M\ast}$, and
\[
g_{A}^{(12)}=-\frac{c_{A_{N}}}{g_{1}}\left(  1-\frac{1}{Z^{2}}\right) \;, \quad
g_{A}^{(34)}=\frac{c_{A_{M}}}{g_{1}}\left(  1-\frac{1}{Z^{2}}\right)
\]
are the `mixed' axial coupling constants of the bare fields $\Psi_{N}$ with
$\Psi_{N\ast}$ and $\Psi_{M}$ with $\Psi_{M\ast}$.

The expressions for the axial coupling constants of the physical fields can be
obtained from the relevant terms of the axial current (\ref{eq:axial current})
after the transformation to parity eigenstates (\ref{eq:Psi phys}) has been
carried out.


\begin{thebibliography}{99}                                                                                               %


\bibitem {gasioro}S.~Gasiorowicz and D.~A.~Geffen,
Rev.\ Mod.\ Phys.\ \textbf{41} (1969) 531.


\bibitem {meissner}
N.~Fettes and U.~-G.~Meissner,
Nucl.\ Phys.\ A \textbf{676} (2000) 311 [hep-ph/0002162];
N.~Fettes and U.~-G.~Meissner,
Nucl.\ Phys.\ A \textbf{693} (2001) 693 [hep-ph/0101030].


\bibitem {jorge}
J.~M.~Alarcon, J.~Martin Camalich and J.~A.~Oller,
Phys.\ Rev.\ D \textbf{85} (2012) 051503 [arXiv:1110.3797 [hep-ph]].


\bibitem {Baru:2010xn}V.~Baru, C.~Hanhart, M.~Hoferichter, B.~Kubis, A.~Nogga
and D.~R.~Phillips,
Phys.\ Lett.\ B \textbf{694} (2011) 473 [arXiv:1003.4444 [nucl-th]].


\bibitem {koch}V.~Koch,
nucl-th/9512029.


\bibitem {lee}B. W. Lee, \textit{Chiral Dynamics}, Gordon and Breach, New
York, 1972.


\bibitem {dmitrasinovic}V.~Dmitrasinovic and F.~Myhrer,
Phys.\ Rev.\ C \textbf{61} (2000) 025205 [hep-ph/9911320].


\bibitem {dmitra2}
V.~Dmitrasinovic, A.~Hosaka and K.~Nagata,
Int.\ J.\ Mod.\ Phys.\ E \textbf{19} (2010) 91 [arXiv:0912.2396 [hep-ph]].


\bibitem {Gallas:2009qp}S.~Gallas, F.~Giacosa and D.~H.~Rischke,
Phys.\ Rev.\ D \textbf{82} (2010) 014004 [arXiv:0907.5084 [hep-ph]].


\bibitem {Gallas:2013ipa}S.~Gallas and F.~Giacosa,
Int.\ J.\ Mod.\ Phys.\ A \textbf{29} (2014) 17, 1450098 [arXiv:1308.4817
[hep-ph]].


\bibitem {denis}D.~Parganlija, F.~Giacosa and D.~H.~Rischke,
Phys.\ Rev.\ D \textbf{82} (2010) 054024 [arXiv:1003.4934 [hep-ph]];
S.~Janowski, D.~Parganlija, F.~Giacosa and D.~H.~Rischke,
Phys.\ Rev.\ D \textbf{84} (2011) 054007 [arXiv:1103.3238 [hep-ph]].


\bibitem {denisnf3}D.~Parganlija, P.~Kovacs, G.~Wolf, F.~Giacosa and
D.~H.~Rischke,
Phys.\ Rev.\ D \textbf{87} (2013) 014011 [arXiv:1208.0585 [hep-ph]].


\bibitem {stani}
S.~Janowski, F.~Giacosa and D.~H.~Rischke,
Phys.\ Rev.\ D \textbf{90} (2014) 11, 114005 [arXiv:1408.4921 [hep-ph]].


\bibitem {nf4}
W.~I.~Eshraim, F.~Giacosa and D.~H.~Rischke,
Eur.\ Phys.\ J.\ A \textbf{51} (2015) 9, 112 [arXiv:1405.5861 [hep-ph]].


\bibitem {ko}P. Ko and S. Rudaz, Phys. Rev. D \textbf{50} (1994) 6877; M.
Urban, M. Buballa and J. Wambach, Nucl. Phys. A \textbf{697} (2002) 338
[hep-ph/0102260].


\bibitem {scalars}
C.~Amsler and F.~E.~Close,
Phys.\ Rev.\ D \textbf{53} (1996) 295 [arXiv:hep-ph/9507326]; 
W.~J.~Lee and D.~Weingarten,
Phys.\ Rev.\ D \textbf{61}, 014015 (2000) [arXiv:hep-lat/9910008];
F.~E.~Close and A.~Kirk,
Eur.\ Phys.\ J.\ C \textbf{21}, 531 (2001) [arXiv:hep-ph/0103173];
F.~Giacosa, T.~Gutsche, V.~E.~Lyubovitskij and A.~Faessler,
Phys.\ Rev.\ D \textbf{72}, 094006 (2005) [arXiv:hep-ph/0509247]; 
F.~Giacosa, T.~Gutsche and A.~Faessler,
Phys. Rev. C \textbf{71}, 025202 (2005) [arXiv:hep-ph/0408085];
H.~Y.~Cheng, C.~K.~Chua and K.~F.~Liu,
Phys.\ Rev.\ D \textbf{74} (2006) 094005 [arXiv:hep-ph/0607206]; 
V.~Mathieu, N.~Kochelev and V.~Vento,
Int.\ J.\ Mod.\ Phys.\ E \textbf{18} (2009) 1 [arXiv:hep-ph/0810.4453];
F.~Br\"{u}nner, D.~Parganlija and A.~Rebhan,
[arXiv:hep-ph/1501.07906].


\bibitem {jaffe}R.~L.~Jaffe,
Phys.\ Rev.\ D \textbf{15} (1977) 267;
R.~L.~Jaffe,
Phys.\ Rev.\ D \textbf{15} (1977) 281;
R.~L.~Jaffe,
Phys.\ Rept.\ \textbf{409} (2005) 1 [Nucl.\ Phys.\ Proc.\ Suppl.\ \textbf{142}
(2005) 343] [arXiv:hep-ph/0409065].

\bibitem{tetraquarks}
L.~Maiani, F.~Piccinini, A.~D.~Polosa and V.~Riquer,
Phys.\ Rev.\ Lett.\ \textbf{93} (2004) 212002 [arXiv:hep-ph/0407017];
F.~Giacosa,
Phys.\ Rev.\ D \textbf{74} (2006) 014028 [arXiv:hep-ph/0605191];
A.~H.~Fariborz, R.~Jora and J.~Schechter,
Phys.\ Rev.\ D \textbf{72} (2005) 034001 [arXiv:hep-ph/0506170];
A.~H.~Fariborz,
Int.\ J.\ Mod.\ Phys.\ A \textbf{19} (2004) 2095 [arXiv:hep-ph/0302133];
M.~Napsuciale and S.~Rodriguez,
Phys.\ Rev.\ D \textbf{70} (2004) 094043;
A.~Heinz, S.~Struber, F.~Giacosa and D.~H.~Rischke,
Phys.\ Rev.\ D \textbf{79} (2009) 037502 [arXiv:hep-ph/08051134].


\bibitem {oller}
J. A. Oller and E. Oset, \emph{Nucl. Phys.} \textbf{A620},
438-456 (1997) [arXiv:hep-ph/9702314]; 
J. A. Oller, E. Oset and J. R.
Pel\'{a}ez, \emph{Phys. Rev. Lett.} \textbf{80}, 3452-3455 (1998)
[arXiv:hep-ph/9803242]; 
J. A. Oller, E. Oset and J. R. Pel\'{a}ez,
\emph{Phys. Rev.} \textbf{D59}, 074001 (1999) [Erratum-ibid. \textbf{D60},
099906 (1999); Erratum-ibid. \textbf{D75}, 099903 (2007)]
[arXiv:hep-ph/9804209].


\bibitem {wolkanowskinew}
T.~Wolkanowski, F.~Giacosa and D.~H.~Rischke,
arXiv:1508.00372 [hep-ph].

\bibitem {pelaezrev}
J.~R.~Pelaez,
arXiv:1510.00653 [hep-ph].

\bibitem {Detar:1988kn}C.~E.~DeTar and T.~Kunihiro,
Phys.\ Rev.\ D \textbf{39} (1989) 2805.


\bibitem {Jido:2001nt}D.~Jido, M.~Oka, and A.~Hosaka,
Prog.\ Theor.\ Phys.\ \textbf{106} (2001) 873 [hep-ph/0110005];
D.~Jido, Y.~Nemoto, M.~Oka and A.~Hosaka,
Nucl.\ Phys.\ A \textbf{671} (2000) 471 [hep-ph/9805306].


\bibitem {ziesche}D.~Zschiesche, L.~Tolos, J.~Schaffner-Bielich and
R.~D.~Pisarski,
Phys.\ Rev.\ C \textbf{75} (2007) 055202 [nucl-th/0608044].


\bibitem {sasakimishustin}C.~Sasaki and I.~Mishustin,
Phys.\ Rev.\ C \textbf{82} (2010) 035204 [arXiv:1005.4811 [hep-ph]].


\bibitem {glozman}L.~Y.~Glozman,
Phys.\ Rept.\ \textbf{444} (2007) 1 [hep-ph/0701081];
T.~D.~Cohen and L.~Y.~Glozman,
Int.\ J.\ Mod.\ Phys.\ A \textbf{17} (2002) 1327 [hep-ph/0201242];
L.~Y.~Glozman,
Phys.\ Lett.\ B \textbf{539} (2002) 257 [hep-ph/0205072].


\bibitem {achim}
A.~Heinz, F.~Giacosa and D.~H.~Rischke,
Nuclear Physics A, Volume 933, January 2015, Pages 34-42 [arXiv:1312.3244
[nucl-th]].



\bibitem {pptopkaon}T.~Rozek
Phys.\ Lett.\ B\textbf{643} (2006) 251; 
Yu.~Valdau
Phys.\ Rev.\ C \textbf{81} (2010) 045208; 
A.~Budzanowski \textit{et al.,}
Phys.\ Lett.\ B\textbf{ 692} (2010) 10; 
G.~Agakishiev
Phys.\ Rev.\ C \textbf{85} (2012) 035203.


\bibitem {pptoppphi}F.~Balestra \textit{et al.,}
Phys.\ Rev.\ C \textbf{63} (2001) 024004.


\bibitem {pptoppetaprime}F.~Balestra \textit{et al.,}
Phys.\ Lett.\ B\textbf{ 491} (2000) 29.


\bibitem {kaonp}S.~Prakhov \textit{et al.,}
Phys.\ Rev.\ C \textbf{80} (2009) 025204.

\bibitem {hyperonstar}
S.~Weissenborn, D.~Chatterjee and J.~Schaffner-Bielich,
Nucl.\ Phys.\ A \textbf{881} (2012) 62 [arXiv:1111.6049 [astro-ph.HE]].


\bibitem {giuseppe}
A.~Drago, A.~Lavagno, G.~Pagliara and D.~Pigato,
Phys.\ Rev.\ C \textbf{90} (2014) 6, 065809 [arXiv:1407.2843 [astro-ph.SR]].


\bibitem {PDG}K.~A.~Olive \textit{et al.} [Particle Data Group
Collaboration],
Chin.\ Phys.\ C \textbf{38} (2014) 090001.

\bibitem {quarkDiquark}
D.~B.~Lichtenberg, W.~Namgung, E.~Predazzi and J.~G.~Wills,
Phys.\ Rev.\ Lett.\ \textbf{48} (1982) 1653.


\bibitem {tqmix}F.~Giacosa,
Phys.\ Rev.\ D \textbf{75} (2007) 054007 [arXiv:hep-ph/0611388].

\bibitem {Takahashi}T.~T.~Takahashi and T.~Kunihiro,
Phys.\ Rev.\ D \textbf{78} (2008) 011503 [arXiv:0801.4707 [hep-lat]];
T.~T.~Takahashi and T.~Kunihiro,
eConf \textbf{C070910} (2007) 297 [Mod.\ Phys.\ Lett.\ A \textbf{23} (2008)
2340] [arXiv:0711.1961 [hep-lat]].





\bibitem {sheng}
C.~S.~An and B.~S.~Zou,
Sci.\ Sin.\ G \textbf{52} (2009) 1452 [arXiv:0910.4452 [nucl-th]].


\bibitem {liu}
B.~C.~Liu and B.~S.~Zou,
Phys.\ Rev.\ Lett.\ \textbf{96}, 042002 (2006) [nucl-th/0503069].


\bibitem {chao}
X.~Cao, J.~J.~Xie, B.~S.~Zou and H.~S.~Xu,
Phys.\ Rev.\ C \textbf{80} (2009) 025203 [arXiv:0905.0260 [nucl-th]].


\bibitem {psgproc}
W.~I.~Eshraim, S.~Janowski, A.~Peters, K.~Neuschwander and F.~Giacosa,
Acta Phys.\ Polon.\ Supp.\ \textbf{5} (2012) 1101 [arXiv:1209.3976 [hep-ph]];
W.~I.~Eshraim, S.~Janowski, F.~Giacosa and D.~H.~Rischke,
Phys.\ Rev.\ D \textbf{87} (2013) 5, 054036 [arXiv:1208.6474 [hep-ph]].

\bibitem {harada}
  H.~Nishihara and M.~Harada,
  Phys.\ Rev.\ D {\bf 92} (2015) 5,  054022
  [arXiv:1506.07956 [hep-ph]].



\bibitem {pagliara}S.~Gallas, F.~Giacosa and G.~Pagliara,
Nucl.\ Phys.\ A \textbf{872} (2011) 13 [arXiv:1105.5003 [hep-ph]].



\bibitem {phiP}
E. P. Venugopal and B. R. Holstein, Phys. Rev. D \textbf{57} (1998) 4397
[arXiv:hep-ph/9710382]; T. Feldmann, P. Kroll and B. Stech, Phys. Rev. D
\textbf{58} (1998) 114006 [arXiv:hep-ph/9802409].


\bibitem {noether}
E. Noether, Gott. Nachr. \textbf{1918}, 235 (1918) [Transp. Theory Statist.
Phys. \textbf{1}, 186 (1971)] [physics/0503066].














\end{thebibliography}
\end{document}